\documentclass[aps,prl,reprint, floatfix,nofootinbib,amsmath,amssymb,groupedaddress]{revtex4-2}
\usepackage{graphicx}
\usepackage{epstopdf}
\usepackage{mathrsfs}
\usepackage{hyperref}

\renewcommand{\vec}[1]{\mathbf{#1}}
\renewcommand{\Im}{\text{Im}}
\renewcommand{\Re}{\text{Re}}
\newcommand{\G}{\mathcal{G}}
\newcommand{\LL}{\mathcal{L}}
\newcommand{\FS}{\mathcal{S}}
\newcommand{\intk}{\int_{\mathrm{BZ}}\!\frac{d^{d}\vec k}{(2\pi)^{d}}\,}
\newcommand{\ints}{\oint_{\FS_{\lambda}}\!\frac{dS_{\vec k}}{(2\pi)^{d}}\,}
\newcommand{\intin}{\int_{\Omega_{\lambda}}\!\frac{d^{d}\vec k}{(2\pi)^{d}}\,}
\newcommand{\intout}{\int_{\Omega^{c}_{\lambda}}\!\frac{d^{d}\vec k}{(2\pi)^{d}}\,}
\newcommand{\dlam}{\frac{d}{d\lambda}}
\newcommand{\D}{D_{\lambda}}
\newcommand{\DS}{D^{\FS}_{\lambda}}

\DeclareMathOperator{\Tr}{Tr}
\newcommand{\kk}{\mathbf{k}}
\newcommand{\BZ}{\mathrm{BZ}}
\newcommand{\id}{\mathbf{1}}
\newcommand{\ik}{\int_{\BZ}\!\frac{d^d\kk}{(2\pi)^d}}
\newcommand{\isnu}{\oint_{S_{\lambda\nu}}\!\frac{dS_{\kk}}{(2\pi)^d}}

\newcommand{\nn}{\hat n_{\lambda\kk}}
\newcommand{\GG}{\hat G_{\lambda\kk}}
\newcommand{\SSig}{\hat\Sigma_{\lambda\kk}}
\newcommand{\Ph}{\hat\Phi_{\lambda\kk}}
\newcommand{\Lam}{\hat\Lambda_{\lambda\kk}}
\newcommand{\Ll}{\mathcal{L}_{\lambda\kk}}
\newcommand{\Dl}{D_\lambda}

\newcommand{\xinu}{\xi_{\lambda\nu\kk}}
\newcommand{\unu}{u_{\lambda\nu\kk}}
\newcommand{\Znu}{Z_{\lambda\nu\kk}}
\newcommand{\vnu}{v_{\lambda\nu\kk}}

\newcommand{\nnu}{n_{\lambda\nu\kk}}

\hypersetup{hidelinks}

\begin{document}

\title{Exact solutions for kinematics across interaction-deformed Fermi surfaces}

\author{Alexander Kruchkov}

\author{F. Duncan M. Haldane}
\affiliation{Department of Physics, Princeton University, Princeton NJ 08544, USA}

\date{\today}

\begin{abstract}
Luttinger's theorem fixes the volume of the Fermi sea of a Fermi liquid, but not how interactions redistribute occupation around its surface. For an interaction switched on and tuned at fixed density, we derive the exact rates of change of the occupation deficit inside the interacting Fermi surface and of the surplus outside it. Each is the weight swept by the deforming surface plus a change of Luttinger's integrand at fixed momentum; the difference of the surface terms is flux of incoherent weight $1 - Z_{\lambda\vec k}$. These identities hold with or without the Luttinger theorem.
\end{abstract}

\maketitle

Luttinger's theorem~\cite{Luttinger1960,LuttingerWard1960} is one of the few exact results for interacting fermions: the volume enclosed by the (interacting) Fermi surface is fixed by the particle density, no matter how strongly the interaction reshapes the surface \cite{AGD1963,Nozieres1964}. This theorem is so profound that it underlies multiple experimental and theoretical measures, including the counting of carriers by photoemission and quantum oscillations~\cite{Damascelli2003,DoironLeyraud2007}.  Deviations from Luttinger theorem signal quantum states beyond the Fermi liquid \cite{Dzyaloshinskii2003,DavePhillipsKane2013}. Modern proofs and reformulations cast the Luttinger theorem as a filling constraint~\cite{Oshikawa2000,ParamekantiVishwanath2004} of Lieb--Schultz--Mattis type~\cite{Lieb1961, Hastings2004}, as an anomaly of the emergent symmetry \cite{ElseThorngrenSenthil2021}, or as a winding number of the full interacting  propagator~\cite{SekiYunoki2017}. All of these fix a single number, the Fermi volume, but do not address kinematics and quantum  flows of quasiparticles and quasiholes upon interaction-induced Fermi surface reshaping.

In this Letter we make this redistribution explicit. Following the ground state as the interaction is switched on at fixed density, we track the occupation missing inside the interacting Fermi surface and the occupation present outside it, Eq.~\eqref{eq:setup_N1}, and derive exact expressions for their rates of change with the interaction parameter strength  $\lambda$. Each rate has two parts, coming separately from Fermi surface and Fermi volume.  These identities are kinematic: \textit{they hold for any causal self-energy that is real at the Fermi level and has a quasiparticle pole on a smooth Fermi surface, whether or not Luttinger's theorem holds}. The generalized identities thus give a momentum-resolved account of the redistribution in which Luttinger's theorem is one condition.

\begin{figure}[t]
\includegraphics[width=0.9 \columnwidth]{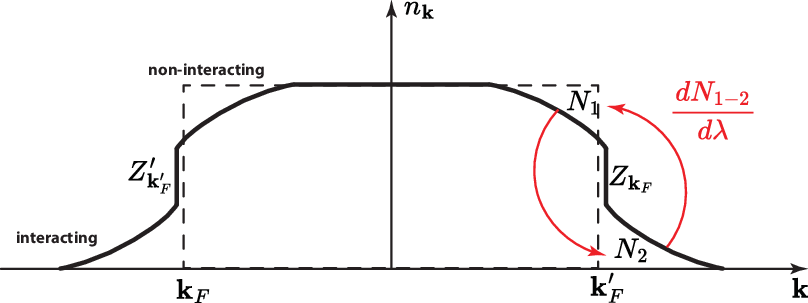}
\caption{Momentum distribution $n_{\lambda\vec k}$ along a line through the Fermi sea (free fermions dashed line, interacting fermions solid line). The interaction moves weight from inside the Fermi surface to outside it: the quasihole deficit $N_{1}$ inside and the quasiparticle surplus $N_{2}$ outside, Eq.~\eqref{eq:setup_N1}.  The distribution keeps a discontinuity at each crossing $\vec k_{F}$ of the surface $\mathcal{S}_{\lambda}(\vec k)$, equal to the quasiparticle weight $Z_{\lambda\vec k_{F}}$ there. }
\label{fig:Fermi_step}
\end{figure}

We consider spinless fermions  on a lattice in $d$ dimensions at $T = 0$, described by the propagator $\G_{\lambda}(\omega,\vec k)$ and the self-energy $\Sigma_{\lambda}(\omega,\vec k)$  describing interactions \cite{AGD1963,Mahan2000}. The momentum distribution $n_{\lambda\vec k}$ (number of particles at given momentum) and the interacting shape of the Fermi surface are fully defined through \textit{interacting} Green's functions $\G_{\lambda}$, as prescribed in the \textit{Formalism} section  below. The interactions push spectral weight from inside the Fermi surface to outside it (Fig.~\ref{fig:Fermi_step}); the missing occupation inside and the excess occupation outside,   \begin{align}
N_{1} \, & = \int_{\substack{\text{inside}}} \frac{d^{d}\vec k}{(2\pi)^{d}} \, (1 - n_{\lambda\vec k}) , 
\\
N_{2} \, &  = \int_{\substack{\text{outside}}} \frac{d^{d}\vec k}{(2\pi)^{d}} \, n_{\lambda\vec k} , \label{eq:setup_N1}
\end{align}
obey $N_{1-2} \equiv N_{1} - N_{2} = V_{F} - N_{\mathrm{tot}}$ by construction, with $V_{F}$ the Fermi volume in dimensionless units, Eq.~\eqref{eq:setup_VF}, and $N_{\mathrm{tot}}$ the density (measure $d^{d}\vec k/(2\pi)^{d}$ throughout). In these notations, Luttinger's theorem simply means $V_{F} = N_{\mathrm{tot}}$, or $N_{1} = N_{2}$, for \textit{any} interaction strength.

We call $N_{1}$ the \textit{quasihole deficit} and $N_{2}$ the \textit{quasiparticle surplus}; they are properties of the ground state, deviations of its occupation from the Fermi step (Fig. 1). Consider further the dimensionless interaction switch $\lambda$, such as the interaction is switched on along a smooth family of Hamiltonians $H_{\lambda}$, $\lambda \in [0,1]$, from the noninteracting case $H_{0}$ to full interacting case $H_{1} = H$.  For notational clarity, we focus on a single-band \textit{exact} interacting Green's function $G_{\lambda}(\omega,\vec k)$.   We define the \textit{interacting Fermi surface} through exact interacting Green's function as $\Phi_{\lambda\vec k} = \G^{-1}_{\lambda}(\mu,\vec k)$, the real function whose zero set is the Fermi surface $\mathcal{S}_{\lambda}(\vec k)$ and whose positive region is the Fermi sea, the region where $\G_{\lambda}(\mu,\vec k) > 0$ as in Refs.~\cite{AGD1963,Dzyaloshinskii2003} ($\mu > \omega_{\vec k}$ for free fermions). As a consequence,  $v_{\lambda\vec k}$ is the normal speed of Fermi surface $\mathcal{S}_{\lambda}(\vec k)$ reshaping, which is positive where the sea recedes.

The switch of the interactions leads not only to the redistribution of the particle-hole densities, but also to the discontinuity of particle momentum distribution at the moving Fermi surface, given by the \textit{quasiparticle weight} $Z_{\lambda\vec k}$ (see Fig.~\ref{fig:Fermi_step}). For $\vec k$ on the Fermi surface, $n^{-}_{\lambda\vec k}$ and $n^{+}_{\lambda\vec k}$ denote the limits of $n_{\lambda\vec k'}$ as $\vec k' \to \vec k$ from inside the sea ($\Phi_{\lambda\vec k'} > 0$) and from outside it ($\Phi_{\lambda\vec k'} < 0$), respectively; their difference is the quasiparticle weight, $n^{-}_{\lambda\vec k} - n^{+}_{\lambda\vec k} = Z_{\lambda\vec k}$ (Migdal discontinuity \cite{Migdal1957},  Fig.~1).   In what follows, the global quantities such as  quasiholes deficit $N_{1}$, quasiparticle surplus $N_{2}$, Fermi volume $V_{F}$, and chemical potential $\mu$ change with the interaction strength $\lambda$ upon  rates $d/d\lambda$.
Since we are dealing with the moving Fermi surface carrying discontinuities (jumps) in physical observables, one needs to use the so-called \textit{convective derivative} known from e.g.  Navier--Stokes equations in hydrodynamics for the moving front, and take into account the shift of  the moving boundary.
We define $D_{\lambda}$ as the derivative along the path at fixed $\vec k$: for a Fermi-level quantity such as $\Phi_{\lambda\vec k}$ or $\Sigma_{\lambda}(\mu,\vec k)$ it is $D_{\lambda} = \partial_{\lambda} + (d\mu/d\lambda)\,\partial_{\omega}$, the second term following the drift of the Fermi level.
\footnote{Formally $\partial_{\lambda} + (d\mu/d\lambda)\partial_{\omega}$ is a convective derivative in the sense of fluid mechanics, $D/Dt = \partial_{t} + \vec u\cdot\nabla$, with the coupling as time, the frequency as position, and the drift of the Fermi level as the velocity; it remains a derivative at fixed momentum. For a differentiable extension of a surface quantity, $D^{\mathcal{S}}_{\lambda} = D_{\lambda} + (d\vec k_{\mathcal{S}}/d\lambda)\cdot\nabla_{\vec k}$ is the \textit{material derivative} following the surface, and Eq.~\eqref{eq:antiflow_jump} is the Reynolds transport theorem.}

\textit{Results}. Our main results are exact expressions for the rates of change of the quasihole deficit and the quasiparticle surplus,
\begin{subequations}
\label{eq:hydro_flows_equal}\label{eq:hydro_flows_equal_N1}%
\renewcommand{\theequation}{\ifnum\value{equation}>2 \the\numexpr\value{parentequation}+1\relax\else\theparentequation\fi\ifodd\value{equation}a\else b\fi}%
\begin{align}
\frac{dN_{1}}{d\lambda} \, &= \, - \!\oint \! \frac{dS_{\vec k}}{(2\pi)^{d}} \, \bigl(1 - Z_{\lambda\vec k} - n^{+}_{\lambda\vec k}\bigr) \, v_{\lambda\vec k}
\label{eq:hydro_flows_equal_N1_surface} \\
&\quad + \int_{\Phi_{\lambda\vec k} > 0} \! \frac{d^{d}\vec k}{(2\pi)^{d}} \, D_{\lambda}\mathcal{L}_{\lambda \vec k} ,
\label{eq:hydro_flows_equal_N1_bulk} \\
\frac{dN_{2}}{d\lambda} \, &= \, + \!\oint \! \frac{dS_{\vec k}}{(2\pi)^{d}} \, n^{+}_{\lambda\vec k} \, v_{\lambda\vec k}
\label{eq:hydro_flows_equal_N2_surface} \\
&\quad - \int_{\Phi_{\lambda\vec k} < 0} \! \frac{d^{d}\vec k}{(2\pi)^{d}} \, D_{\lambda}\mathcal{L}_{\lambda \vec k} .
\label{eq:hydro_flows_equal_N2_bulk}
\end{align}
\end{subequations}%
\refstepcounter{equation}\label{eq:hydro_flows_equal_N2}%
Each population changes in two ways, through the weight swept by the moving surface (the surface terms, Eqs.~\eqref{eq:hydro_flows_equal_N1_surface} and \eqref{eq:hydro_flows_equal_N2_surface}) and through the redistribution behind it (the bulk terms, Eqs.~\eqref{eq:hydro_flows_equal_N1_bulk} and \eqref{eq:hydro_flows_equal_N2_bulk}). The two surface terms are  not equal: they share the flux of the surplus, $\oint n^{+}_{\lambda\vec k}v_{\lambda\vec k}$, which cancels in the difference of the flows, while the quasihole flow alone carries the flux of the interaction-induced \textit{incoherent weight} $\oint(1 - Z_{\lambda\vec k})v_{\lambda\vec k}$. Behind the surface, occupation is redistributed at fixed momentum at the rate $-D_{\lambda}\mathcal{L}_{\lambda\vec k}$: Luttinger's integrand  $\mathcal{L}_{\lambda \vec k}$ (defined in Eq.~\eqref{eq:antiflow_IL}) is the density of displaced spectral weight, and its rate is the local rate of the redistribution. This is the first line of the rate identities of the momentum distribution, of Luttinger's integral, and of the residue,

\begin{align}
D_{\lambda} n_{\lambda\vec k} \, &= \, -\,D_{\lambda}\mathcal{L}_{\lambda\vec k} , \qquad \vec k \notin \mathcal{S}_{\lambda} ,
\notag \\
D^{\mathcal{S}}_{\lambda}\bigl[n_{\lambda\vec k}\bigr] \, &  = \, D^{\mathcal{S}}_{\lambda} Z_{\lambda\vec k} ,
\ \ \ \  \qquad \vec k \in \mathcal{S}_{\lambda} ,
\label{eq:antiflow_rates_Z}
\end{align}
Equations~\eqref{eq:hydro_flows_equal_N1}--\eqref{eq:antiflow_rates_Z} are kinematic identities: they \textit{hold for any quasiparticle-like self-energy, whether or not it derives from a Luttinger--Ward functional, and whether or not Luttinger's theorem holds.}

In fact, Luttinger theorem is a special case of Eqs.~\eqref{eq:hydro_flows_equal_N1} and \eqref{eq:hydro_flows_equal_N2} when quasihole deficit cancels with quasiparticle surplus.  When Luttinger's theorem is valid along the path, that is, when $N_{1} = N_{2}$ at every $\lambda$, Eq.~\eqref{eq:LW_real} below: the two rates are equal, with the common value
\begin{align}
\frac{dN_{1}}{d\lambda} \, = \, \frac{dN_{2}}{d\lambda} \, = \, \frac12 \, \frac{d}{d\lambda} \int_{\mathrm{BZ}} \! \frac{d^{d}\vec k}{(2\pi)^{d}} \, \bigl| \mathcal{L}_{\lambda \vec k} \bigr| ,
\label{eq:hydro_flows_L1}
\end{align}
the two populations change at equal rates, half the rate of the total displaced weight $\int|\mathcal{L}_{\lambda\vec k}|$.

 At fixed particle density, whether or not the theorem holds, the bulk redistribution and the surface flux assumes balance,

\begin{align}
\int_{\mathrm{BZ}} \! \frac{d^{d}\vec k}{(2\pi)^{d}} \, D_{\lambda} n_{\lambda \vec k} \, &= \, \oint \! \frac{dS_{\vec k}}{(2\pi)^{d}} \, Z_{\lambda\vec k} \, v_{\lambda\vec k} . \label{eq:hydro_number_flux}
\end{align}
Equation~\eqref{eq:hydro_number_flux} is a statement of particle-number conservation in the case of interaction-deforming Fermi surface. The momenta released from the sea by the receding surface ($v_{\lambda\vec k} > 0$) lose the  weight $Z_{\lambda\vec k}$, and those captured by the advancing surface gain it. At fixed particle number, the swept coherent weight reappears as occupation gained at fixed momentum in the bulk.

\textit{Formalism.---}The momentum distribution at coupling $\lambda$ is the equal-time limit of the Matsubara propagator, equivalently the occupied part of the spectral function~\cite{AGD1963,Mahan2000},
\begin{align}
n_{\lambda\vec k} \, &= \, \frac{1}{\beta}\sum_{\omega_{n}} \G_{\lambda}(\mu + i\omega_{n},\vec k)\,e^{i\omega_{n}0^{+}}
\label{eq:setup_matsubara}
\end{align}
with $\omega_{n} = (2n+1)\pi / \beta$, $\beta = 1/T$; the shift by $\mu$ arises because frequencies are measured from the zero of energy, as in Eq.~\eqref{eq:setup_Ginv}.  With the spectral function $A_{\lambda} = -\frac{1}{\pi}\Im\,\G^{R}_{\lambda} \ge 0$, at $T = 0$, the limit used throughout, Eq.~\eqref{eq:setup_matsubara} becomes
\begin{align}
n_{\lambda\vec k} \, = \int_{-\infty}^{\mu} \! d\omega \, A_{\lambda}(\omega,\vec k) , \label{eq:setup_nT0}
\end{align}

We highlight that we consider the most generic interacting system, which needs to be solved self-consistently for any interaction strength, and where interactions enter through self-energy operator $\Sigma_{\lambda}(\omega,\vec k) $ parametrized by interaction switch $\lambda$,
\begin{align}
	\G^{-1}_{\lambda}(\omega,\vec k) = \omega - \omega_{\vec k} - \Sigma_{\lambda}(\omega,\vec k) . \label{eq:setup_Ginv}
	\end{align}
 While this type of equation is in reality complicated, using it has an added benefit that we can define \textit{exact shape of the interacting Fermi surface} $\mathcal{S}_{\lambda} (\vec k) $. In fact, the exact interacting Fermi surface is where the inverse interacting propagator $\G^{-1}_{\lambda}$  vanishes at the Fermi level \cite{Luttinger1960,Dzyaloshinskii2003},
\begin{align}
\Phi_{\lambda\vec k} \, &\equiv \, \G^{-1}_{\lambda}(\mu,\vec k) \, = \, \mu - \omega_{\vec k} - \Sigma_{\lambda}(\mu,\vec k) ,
\notag \\
\mathcal{S}_{\lambda} (\vec k) \, &= \, \{\vec k : \Phi_{\lambda\vec k} = 0\} ,
\label{eq:flow_phi}
\end{align}
 $\Phi_{\lambda\vec k}$ changes with the coupling through the self-energy at fixed frequency, through the frequency argument $\mu(\lambda)$ of the self-energy, and through the explicit $\mu(\lambda)$, so that at fixed momentum $D_{\lambda}\Phi_{\lambda\vec k} = (d\mu/d\lambda)[1 - \partial_{\omega}\Sigma_{\lambda}(\mu,\vec k)] - \partial_{\lambda}\Sigma_{\lambda}(\mu,\vec k)$.

\textit{Quasiparticles}. While Eq.~\eqref{eq:setup_Ginv} is the most generic form of the interacting propagator, it is useful to highlight its structure in the presence of interaction-induced quasiparticles.  Near the Fermi surface,  the (retarded) interacting propagator $\mathcal G_{\lambda}$  develops a \textit{quasiparticle pole} \cite{PinesNozieres1966,Mahan2000}
\begin{align}
\G^{R}_{\lambda}(\omega,\vec k)  =   \frac{Z_{\lambda\vec k}}{\omega - \tilde\omega_{\lambda\vec k} + i \gamma}  + \text{Regular}, \label{eq:setup_pole}
\end{align}
where $\tilde\omega_{\lambda\vec k} \, = \, \omega_{\vec k} + \Re\Sigma^{R}_{\lambda}(\tilde\omega_{\lambda\vec k},\vec k)$ is  renormalized quasiparticle dispersion, and $Z_{\lambda\vec k}$, a discontinuity illustrated in Fig.~\ref{fig:Fermi_step}, is the quasiparticle residue, defined by interactions
\begin{align}
Z_{\lambda\vec k}^{-1} \, & = \, 1 - \partial_{\omega}\Re\Sigma^{R}_{\lambda}(\tilde\omega_{\lambda\vec k},\vec k) .
\label{eq:setup_Z}
\end{align}%
Analyzing now the definition of the interacting Fermi surface, Eq.~\eqref{eq:flow_phi}, we understand that $\mathcal{S}_{\lambda}(\vec k)$ is also the locus $\tilde\omega_{\lambda\vec k} = \mu$, where $\Phi_{\lambda\vec k} \simeq (\mu - \tilde\omega_{\lambda\vec k})/Z_{\lambda\vec k}$, and the momentum distribution \eqref{eq:setup_nT0} drops by the residue across it, $n^{-}_{\lambda\vec k} - n^{+}_{\lambda\vec k} = Z_{\lambda\vec k}$ (as illustrated in Fig.~\ref{fig:Fermi_step}).

We further define the Fermi volume and the total density of an \textit{interacting} system as  
\begin{align}
V_{F} \, = \int_{\Phi_{\lambda\vec k} > 0} \frac{d^{d}\vec k}{(2\pi)^{d}}, \qquad
N_{\mathrm{tot}} \, = \int_{\mathrm{BZ}} \frac{d^{d}\vec k}{(2\pi)^{d}} \, n_{\lambda\vec k} ,
\label{eq:setup_VF}
\end{align}
hence  Eq.~\eqref{eq:setup_N1} gives $N_{1} - N_{2} = V_{F} - N_{\mathrm{tot}}$.  Luttinger's theorem, in our notations the simple statement $V_{F} = N_{\mathrm{tot}}$~\cite{Luttinger1960}, is invoked only where stated below, so the rate of $V_{F}$ is kept explicit. For an arbitrary interaction, $V_F$ may or may not be equal to $N_{\mathrm{tot}}$. 

\textit{Derivation of the flows.} To calculate the flow of quasiholes and quasiparticles across the surface, we now take into account that both the occupations and the regions in Eq.~\eqref{eq:setup_N1} depend on interaction parameter $\lambda$,
\begin{align}
\frac{dN_{1}}{d\lambda} \, &= \, \frac{d}{d\lambda} \int_{\Phi_{\lambda\vec k} > 0} \frac{d^{d}\vec k}{(2\pi)^{d}} \, (1 - n_{\lambda \vec k}) , \label{eq:setup_dN1} \\
\frac{dN_{2}}{d\lambda} \, &= \, \frac{d}{d\lambda} \int_{\Phi_{\lambda\vec k} < 0} \frac{d^{d}\vec k}{(2\pi)^{d}} \, n_{\lambda \vec k} , \label{eq:setup_dN2}
\end{align}
where the ``$1$'' in the integrand of Eq. \eqref{eq:setup_dN1} integrates to the Fermi volume $V_F$  defined in Eq.  \eqref{eq:setup_VF}.

Note that Eqs.~\eqref{eq:setup_dN1} and \eqref{eq:setup_dN2} are exact, however not so much comfortable to work with, as they involve microscopic-dependent shapes of ``inside'' and ``outside'' of Fermi surface regions within Brillouin Zone (BZ). It would be thus much more comfortable to redefine Eqs.~\eqref{eq:setup_dN1} and \eqref{eq:setup_dN2} on the entire BZ torus, which can be done by using  step functions $\Theta(x)$,
\begin{align}
\frac{dN_{1}}{d\lambda} \, &= \, \frac{d}{d\lambda} \int_{\mathrm{BZ}} \frac{d^{d}\vec k}{(2\pi)^{d}} \, \Theta(\Phi_{\lambda\vec k}) \, (1 - n_{\lambda \vec k}) , \label{eq:setup_dN1_theta} \\
\frac{dN_{2}}{d\lambda} \, &= \, \frac{d}{d\lambda} \int_{\mathrm{BZ}} \frac{d^{d}\vec k}{(2\pi)^{d}} \, \Theta(-\Phi_{\lambda\vec k}) \, n_{\lambda \vec k} .  \label{eq:setup_dN2_theta}
\end{align}
The advantage of this definition is that all the  $\lambda$-dependence of the ``inside'' and ``outside'' domains is now sitting in $\Theta(\pm\Phi_{\lambda\vec k})$. For any quantity $F_{\lambda\vec k}$ with a Migdal-type discontinuity on the moving Fermi surface, the rate of change of its BZ integral is now given by
\begin{align}
\frac{d}{d\lambda} \int_{\mathrm{BZ}} \frac{d^{d}\vec k}{(2\pi)^{d}} \, F_{\lambda \vec k} \, = \, & \int_{\mathrm{BZ}} \frac{d^{d}\vec k}{(2\pi)^{d}} \, D_{\lambda}F_{\lambda \vec k}
\notag \\
& - \oint \! \frac{dS_{\vec k}}{(2\pi)^{d}} \, [F_{\lambda\vec k}] \, v_{\lambda\vec k} ,
\label{eq:antiflow_jump}
\end{align}
the Reynolds transport theorem for a field with a moving discontinuity~\cite{Slattery1990, Truesdell1960, Aris1962}. The first term is the redistribution of $F_{\lambda\vec k}$ at fixed momentum behind the surface; the second is the discontinuity $[F_{\lambda\vec k}] = F^{-}_{\lambda\vec k} - F^{+}_{\lambda\vec k}$ (inside minus outside, the limits defined as for $n^{\mp}_{\lambda\vec k}$) swept by the moving surface, $-v_{\lambda\vec k}\,dS_{\vec k}/(2\pi)^{d}$ being the volume gained by the interacting Fermi sea through the element $dS_{\vec k}$. For $F_{\lambda \vec k} = n_{\lambda \vec k}$ the discontinuity is the residue, $[n_{\lambda \vec k}] = Z_{\lambda \vec k}$. Since $dN_{\mathrm{tot}}/d\lambda = 0$ at fixed density,  Eq.~\eqref{eq:antiflow_jump} gives $\int_{\mathrm{BZ}} d^{d}\vec k\, D_{\lambda}n_{\lambda\vec k}/(2\pi)^{d} = \oint dS_{\vec k}\, Z_{\lambda\vec k}\, v_{\lambda\vec k}/(2\pi)^{d}$, Eq.~\eqref{eq:hydro_number_flux}.

The two terms of Eq.~\eqref{eq:antiflow_jump}, for the sea and for the exterior, follow from Eqs.~\eqref{eq:setup_dN1_theta} and \eqref{eq:setup_dN2_theta}: the rate of a step-weighted integrand has a surface part and a bulk part, which for $F_{\lambda\vec k} = n_{\lambda\vec k}$ read
\begin{align}
\frac{d}{d\lambda}\bigl[\Theta(\pm\Phi_{\lambda\vec k})\,n_{\lambda\vec k}\bigr] \, = \, & \pm\,\delta(\Phi_{\lambda\vec k})\,D_{\lambda}\Phi_{\lambda\vec k}\,n^{\mp}_{\lambda\vec k}
\notag \\
& + \, \Theta(\pm\Phi_{\lambda\vec k})\,D_{\lambda}n_{\lambda\vec k} .
\label{eq:flow_lift}
\end{align}%
Here $d/d\lambda$ is the derivative with respect to $\lambda$ at fixed $\vec k$ taken in the sense of distributions in $\vec k$, the derivative that commutes with the zone integral in Eqs.~\eqref{eq:setup_dN1_theta}--\eqref{eq:antiflow_jump}: unlike $D_{\lambda}$, the classical rate on either side of the surface, it also differentiates the moving step, and the resulting $\delta$ function carries the swept discontinuity. The left-hand side of \eqref{eq:flow_lift} is thus the full rate of the region-restricted occupation, swept discontinuity included; on the right, $D_{\lambda}n_{\lambda\vec k}$ is the bulk rate within the region and $n^{\mp}_{\lambda\vec k}$ the value of $n_{\lambda\vec k}$ at the surface on the side of the region ($n^{-}_{\lambda\vec k}$ for the sea, $n^{+}_{\lambda\vec k}$ for the exterior); for any other quantity $F_{\lambda\vec k}$ that jumps on the surface, $n$ is replaced by $F$, and $F_{\lambda\vec k} = 1$ gives the rate of the Fermi volume. The $\delta$ function turns the surface part into an integral over $\mathcal{S}_{\lambda}(\vec k)$ with the element $dS_{\vec k}/|\nabla_{\vec k}\Phi_{\lambda\vec k}|$, so that it equals $\mp\oint n^{\mp}_{\lambda\vec k}\,v_{\lambda\vec k}\,dS_{\vec k}/(2\pi)^{d}$,
 with the signed normal speed $v_{\lambda\vec k} \equiv -D_{\lambda}\Phi_{\lambda\vec k}/|\nabla_{\vec k}\Phi_{\lambda\vec k}|$ along the inward normal $\hat n = \nabla_{\vec k}\Phi_{\lambda\vec k}/|\nabla_{\vec k}\Phi_{\lambda\vec k}|$.  $N_{1}$ changes by $(1 - n^{-}_{\lambda\vec k})$ times the volume gained by the sea and $N_{2}$ by $-n^{+}_{\lambda\vec k}$ times it.  Therefore, Eqs.~\eqref{eq:antiflow_jump} and \eqref{eq:flow_lift} give
\begin{align}
\frac{dN_{1}}{d\lambda} \, &= \, - \!\oint \! \frac{dS_{\vec k}}{(2\pi)^{d}} \, (1 - n^{-}_{\lambda\vec k})\, v_{\lambda\vec k} \, - \int_{\Phi_{\lambda\vec k} > 0} \! \frac{d^{d}\vec k}{(2\pi)^{d}} \, D_{\lambda} n_{\lambda \vec k} ,
\label{eq:flow_dN1} \\
\frac{dN_{2}}{d\lambda} \, &= \, + \!\oint \! \frac{dS_{\vec k}}{(2\pi)^{d}} \, n^{+}_{\lambda\vec k}\, v_{\lambda\vec k} \, + \int_{\Phi_{\lambda\vec k} < 0} \! \frac{d^{d}\vec k}{(2\pi)^{d}} \, D_{\lambda} n_{\lambda \vec k} .
\label{eq:flow_dN2}
\end{align}%
 With Migdal's relation for the discontinuity of momentum distribution at the interacting Fermi surface $1 - n^{-}_{\lambda\vec k} = (1 - Z_{\lambda\vec k}) - n^{+}_{\lambda\vec k}$ in the surface term of \eqref{eq:flow_dN1}, and with $D_{\lambda}n_{\lambda\vec k} = -D_{\lambda}\mathcal{L}_{\lambda\vec k}$ off the surface, the first line of Eq. \eqref{eq:antiflow_rates_Z} (derived below), in the bulk terms, these are Eqs.~\eqref{eq:hydro_flows_equal_N1} and \eqref{eq:hydro_flows_equal_N2}.

\textit{Motion of the Fermi surface.---}On the surface the element $dS_{\vec k}/|\nabla_{\vec k}\Phi_{\lambda\vec k}| = Z_{\lambda\vec k}\,dS_{\vec k}/|\nabla_{\vec k}\tilde\omega_{\lambda\vec k}|$, and the normal speed of  the Fermi-surface shift takes the physical form
\begin{align}
v_{\lambda\vec k} \, = \, - \frac{D_{\lambda}\Phi_{\lambda \vec k}}{|\nabla_{\vec k} \Phi_{\lambda \vec k} |}
\, = \, \frac{Z_{\lambda\vec k}\, \partial_{\lambda}\Sigma_{\lambda}(\mu,\vec k) \, - \, d\mu/d\lambda}{|\nabla_{\vec k}\tilde\omega_{\lambda\vec k}|} ,
\label{eq:hydro_vn_physical}
\end{align}
the second form following from \eqref{eq:setup_Z} differentiated in $\vec k$ and along the path. Hence,  the interacting Fermi surface moves at the rate at which the quasiparticle level drifts relative to the chemical potential, divided by the renormalized Fermi  velocity. The Fermi volume changes by the net normal flux, $dV_{F}/d\lambda = -\oint v_{\lambda\vec k}\,dS_{\vec k}/(2\pi)^{d}$. At fixed density Luttinger's theorem therefore says that the net flux vanishes, so that the Fermi sea deforms at constant volume, and solving $\oint v_{\lambda\vec k}\,dS_{\vec k} = 0$ for the drift of the chemical potential gives
\begin{align}
\frac{d\mu}{d\lambda} \, = \, \bigl\langle Z_{\lambda\vec k}\,\partial_{\lambda}\Sigma_{\lambda}(\mu,\vec k) \bigr\rangle_{\mathcal{S}_{\lambda}} ,
\label{eq:hydro_mudot}
\end{align}
the average over the surface weighted with the quasiparticle density of states, $\langle X\rangle_{\mathcal{S}_{\lambda}} = \oint X\,dS_{\vec k}/|\nabla_{\vec k}\tilde\omega_{\lambda\vec k}| \big/ \oint dS_{\vec k}/|\nabla_{\vec k}\tilde\omega_{\lambda\vec k}|$.  Therefore, the chemical potential absorbs the average drift of the quasiparticle levels on the Fermi surface.

\textit{Luttinger's terms.---} Off the Fermi surface $\mathcal S_{\lambda}(\vec k)$, the change of $n_{\lambda\vec k}$ at fixed momentum is governed by Luttinger-type integrand \cite{Luttinger1960,LuttingerWard1960},
\begin{align}
\mathcal{L}_{\lambda \vec k} \, \equiv \, \frac{1}{\pi}\Im \int_{-\infty}^{\mu} \! d\omega \, \G^{R}_{\lambda}(\omega, \vec k) \, \partial_{\omega} \Sigma^{R}_{\lambda}(\omega, \vec k) .
\label{eq:antiflow_IL}
\end{align}
 Writing $\G^{R}_{\lambda} = \partial_{\omega}\ln\G^{R\,-1}_{\lambda} + \G^{R}_{\lambda}\,\partial_{\omega}\Sigma^{R}_{\lambda}$ in Eq.~\eqref{eq:setup_nT0}, the total derivative integrates to the phase of the inverse propagator,
\begin{align}
	n_{\lambda\vec k} = -\pi^{-1}\bigl[\arg\G^{-1}_{\lambda}(\omega + i0^{+},\vec k)\bigr]_{\omega = -\infty}^{\omega = \mu} - \mathcal{L}_{\lambda\vec k} .
\label{eq:antiflow_phase}
\end{align}
 The branch of the logarithm is fixed by causality and by its high-frequency normalization: $\Im\Sigma^{R}_{\lambda} \le 0$ gives $\Im\G^{-1}_{\lambda}(\omega + i0^{+}) = 0^{+} - \Im\Sigma^{R}_{\lambda}(\omega) > 0$ at every real $\omega$, so the inverse propagator never leaves the upper half-plane and its phase, continuous in $\omega$, stays in the range $[0,\pi]$.  Furthermore, for any interacting system, its ultraviolet (large-frequency) behavior is  $\G_{\lambda}(\omega ) \sim  1/ \omega $ as $\omega \to -\infty$, since $\Sigma^{R}_{\lambda}$ is bounded, hence the phase fixes there at $\pi$. At $\omega = \mu$, $\G^{-1}_{\lambda}(\mu ,\vec k) = \Phi_{\lambda\vec k} $ is real and finite, with phase $0$ inside the sea ($\Phi_{\lambda\vec k} > 0$) and $\pi$ outside ($\Phi_{\lambda\vec k} < 0$). Therefore, off the Fermi surface $\mathcal{S}_{\lambda}(\vec k)$, Eq.~\eqref{eq:antiflow_phase} can be written as
\begin{align}
n_{\lambda\vec k} \, = \, \Theta(\Phi_{\lambda\vec k}) \, - \, \mathcal{L}_{\lambda\vec k} , \qquad \vec k \notin \mathcal{S}_{\lambda} ,
\label{eq:antiflow_pointwise}
\end{align}
Eq.~\eqref{eq:antiflow_pointwise} in its differential form is given by the first line of Eq.~\eqref{eq:antiflow_rates_Z}.
We note that Eq.~\eqref{eq:antiflow_pointwise}  uses only causality, the boundedness of $\Sigma^{R}_{\lambda}$ at large $|\omega|$, and the reality of $\Sigma_{\lambda}(\mu,\vec k)$. Together with the transport equations \eqref{eq:antiflow_jump}, and differentiated at fixed momentum off the surface, where $\Theta(\Phi_{\lambda\vec k})$ is constant, it is the first line of Eq.~\eqref{eq:antiflow_rates_Z}, which turns the bulk terms of Eqs.~\eqref{eq:flow_dN1} and \eqref{eq:flow_dN2} into rates of $\mathcal{L}_{\lambda\vec k}$ and so gives the flows \eqref{eq:hydro_flows_equal_N1} and \eqref{eq:hydro_flows_equal_N2}; followed along the moving surface, where its discontinuities are $[n_{\lambda\vec k}] = Z_{\lambda\vec k}$ and $[\mathcal{L}_{\lambda\vec k}] = 1 - Z_{\lambda\vec k}$, it is the second line; and integrated over the two regions, where $\mathcal{L}_{\lambda\vec k} = 1 - n_{\lambda\vec k} \ge 0$ inside and $\mathcal{L}_{\lambda\vec k} = -n_{\lambda\vec k} \le 0$ outside, it gives
\begin{align}
N_{1-2} \equiv N_{1} - N_{2} \, &= \int_{\mathrm{BZ}} \frac{d^{d}\vec k}{(2\pi)^{d}} \, \mathcal{L}_{\lambda \vec k} , \label{eq:antiflow_N12_is_L} \\
N_{1+2} \equiv N_{1} + N_{2} \, &= \int_{\mathrm{BZ}} \frac{d^{d}\vec k}{(2\pi)^{d}} \, \bigl|\mathcal{L}_{\lambda \vec k}\bigr| . \label{eq:antiflow_L1norm}
\end{align}
Equations~\eqref{eq:antiflow_N12_is_L} and \eqref{eq:antiflow_L1norm} hold at any interaction strength $\lambda$ and carry no surface term.

By Eq.~\eqref{eq:antiflow_N12_is_L}, Luttinger's theorem, $N_{1} = N_{2}$, is equivalent to the statement~\cite{LuttingerWard1960}
\begin{align}
\int_{\mathrm{BZ}} \frac{d^{d}\vec k}{(2\pi)^{d}} \, \mathcal{L}_{\lambda \vec k} = 0 \qquad \text{for every } \lambda \in [0,1] ,
\label{eq:LW_real}
\end{align}
which is the conventional path to prove Luttinger theorem through invoking Luttinger-Ward functionals.

\textit{Surface--bulk balance.} We can further rewrite the bulk redistribution as a suface integral. Indeed, using Eq.  \eqref{eq:antiflow_jump} applied to momentum-resolved Luttinger integrand $\mathcal{L}_{\lambda\vec k}$, whose discontinuity is the incoherent weight $1 - Z_{\lambda\vec k}$, together with Eq.~\eqref{eq:LW_real} at every $\lambda$, gives identity
\begin{align}
  \int_{\mathrm{BZ}} \frac{d^{d}\vec k}{(2\pi)^{d}} \, D_{\lambda}\mathcal{L}_{\lambda \vec k}
 = \oint \! \frac{dS_{\vec k}}{(2\pi)^{d}} \, (1 - Z_{\lambda\vec k}) \, v_{\lambda\vec k} .
\label{eq:antiflow_zero}
\end{align}
 Combined with the first line of Eq.~\eqref{eq:antiflow_rates_Z} and with $\oint v_{\lambda\vec k}\,dS_{\vec k} = 0$, this gives  Eq.~\eqref{eq:hydro_number_flux}. In general the moving surface sweeps a non-zero flux of incoherent weight that the bulk must redistribute. The balance follows from particle-number conservation, Migdal's discontinuity, and the volume constraint of the Luttinger--Ward identity. In general, both sides of  Eq.~\eqref{eq:antiflow_zero} are nonzero upon switching on, or modifying  interactions. Their difference is $d(N_{1} - N_{2})/d\lambda$, by Eq.~\eqref{eq:antiflow_jump} for $F_{\lambda\vec k} = \mathcal{L}_{\lambda\vec k}$ and Eq.~\eqref{eq:antiflow_N12_is_L}, so they are equal when Luttinger's theorem holds.

\

\textit{Discussion.---} The derived kinematic identities~\eqref{eq:hydro_flows_equal_N1} and \eqref{eq:hydro_flows_equal_N2} and their consequences \eqref{eq:antiflow_rates_Z},  \eqref{eq:hydro_number_flux}  describe how the quasiparticles and quasiholes are redistributed through the interacting Fermi surface as the interaction is switched on, whether or not Luttinger's theorem holds.  The theorem is the equality of the two rates for $N_1$ and $N_2$ at every interaction strength, which may or may not hold. When it holds, the Fermi surface deforms at constant volume, the chemical potential follows the drift of the quasiparticle levels weighted by the quasiparticle density of states, Eq.~\eqref{eq:hydro_mudot}, and the zone-integrated change of Luttinger's integrand equals the flux of incoherent weight swept by the surface, Eq.~\eqref{eq:antiflow_zero}. Quasiparticle residues define occupation flows \eqref{eq:hydro_number_flux}. 

Our results may be useful in two ways. On the theory side, our formalism provides a framework for studying how Fermi-surface observables evolve with interactions, and could be used, for example,  to address the Fermi surface contributions of intrinsic anomalous Hall response \cite{Haldane2004} in the fully interacting case. On the experimental side, \textit{in situ} twisting of moiré heterostructures \cite{Inbar2023} enables continuous tuning of the interaction parameter $\lambda$ without changing materials and samples; measure of interaction-reshaped Fermi surfaces and flows could help extract valuable momentum-resolved self-energy signatures.

\bibliographystyle{apsrev4-1}
\bibliography{Ref}

@article{Luttinger1960,
  author  = {Luttinger, J. M.},
  title   = {Fermi Surface and Some Simple Equilibrium Properties of a System of Interacting Fermions},
  journal = {Phys. Rev.},
  volume  = {119},
  pages   = {1153},
  year    = {1960},
  doi     = {10.1103/PhysRev.119.1153},
  url     = {https://doi.org/10.1103/PhysRev.119.1153}
}

@article{LuttingerWard1960,
  author  = {Luttinger, J. M. and Ward, J. C.},
  title   = {Ground-State Energy of a Many-Fermion System. {II}},
  journal = {Phys. Rev.},
  volume  = {118},
  pages   = {1417},
  year    = {1960},
  doi     = {10.1103/PhysRev.118.1417},
  url     = {https://doi.org/10.1103/PhysRev.118.1417}
}

@book{AGD1963,
  author    = {Abrikosov, A. A. and Gor'kov, L. P. and Dzyaloshinski, I. E.},
  title     = {Methods of Quantum Field Theory in Statistical Physics},
  publisher = {Prentice-Hall},
  address   = {Englewood Cliffs, NJ},
  year      = {1963}
}

@book{Nozieres1964,
  author    = {Nozi{\`e}res, P.},
  title     = {Theory of Interacting {F}ermi Systems},
  publisher = {Benjamin},
  address   = {New York},
  year      = {1964}
}

@article{Damascelli2003,
  author  = {Damascelli, Andrea and Hussain, Zahid and Shen, Zhi-Xun},
  title   = {Angle-resolved photoemission studies of the cuprate superconductors},
  journal = {Rev. Mod. Phys.},
  volume  = {75},
  pages   = {473--541},
  year    = {2003},
  doi     = {10.1103/RevModPhys.75.473}
}

@article{DoironLeyraud2007,
  author  = {Doiron-Leyraud, Nicolas and Proust, Cyril and LeBoeuf, David and Levallois, Julien and Bonnemaison, Jean-Baptiste and Liang, Ruixing and Bonn, D. A. and Hardy, W. N. and Taillefer, Louis},
  title   = {Quantum oscillations and the {F}ermi surface in an underdoped high-{$T_c$} superconductor},
  journal = {Nature},
  volume  = {447},
  pages   = {565--568},
  year    = {2007},
  doi     = {10.1038/nature05872}
}

@article{Oshikawa2000,
  author  = {Oshikawa, Masaki},
  title   = {Topological Approach to {L}uttinger's Theorem and the {F}ermi Surface of a {K}ondo Lattice},
  journal = {Phys. Rev. Lett.},
  volume  = {84},
  pages   = {3370},
  year    = {2000},
  doi     = {10.1103/PhysRevLett.84.3370},
  url     = {https://doi.org/10.1103/PhysRevLett.84.3370}
}

@article{Dzyaloshinskii2003,
  author  = {Dzyaloshinskii, I.},
  title   = {Some consequences of the {L}uttinger theorem: The {L}uttinger surfaces in non-{F}ermi liquids and {M}ott insulators},
  journal = {Phys. Rev. B},
  volume  = {68},
  pages   = {085113},
  year    = {2003},
  doi     = {10.1103/PhysRevB.68.085113},
  url     = {https://doi.org/10.1103/PhysRevB.68.085113}
}

@article{DavePhillipsKane2013,
  author  = {Dave, Kiaran B. and Phillips, Philip W. and Kane, Charles L.},
  title   = {Absence of {L}uttinger's theorem due to zeros in the single-particle {G}reen function},
  journal = {Phys. Rev. Lett.},
  volume  = {110},
  pages   = {090403},
  year    = {2013},
  doi     = {10.1103/PhysRevLett.110.090403}
}

@article{ParamekantiVishwanath2004,
  author  = {Paramekanti, Arun and Vishwanath, Ashvin},
  title   = {Extending {L}uttinger's theorem to {$Z_2$} fractionalized phases of matter},
  journal = {Phys. Rev. B},
  volume  = {70},
  pages   = {245118},
  year    = {2004},
  doi     = {10.1103/PhysRevB.70.245118}
}

@article{Hastings2004,
  author  = {Hastings, M. B.},
  title   = {{L}ieb-{S}chultz-{M}attis in higher dimensions},
  journal = {Phys. Rev. B},
  volume  = {69},
  pages   = {104431},
  year    = {2004},
  doi     = {10.1103/PhysRevB.69.104431}
}

@article{ElseThorngrenSenthil2021,
  author  = {Else, Dominic V. and Thorngren, Ryan and Senthil, T.},
  title   = {Non-{F}ermi liquids as ersatz {F}ermi liquids: General constraints on compressible metals},
  journal = {Phys. Rev. X},
  volume  = {11},
  pages   = {021005},
  year    = {2021},
  doi     = {10.1103/PhysRevX.11.021005}
}

@article{SekiYunoki2017,
  author  = {Seki, Kazuhiro and Yunoki, Seiji},
  title   = {Topological interpretation of the {L}uttinger theorem},
  journal = {Phys. Rev. B},
  volume  = {96},
  pages   = {085124},
  year    = {2017},
  doi     = {10.1103/PhysRevB.96.085124},
  url     = {https://doi.org/10.1103/PhysRevB.96.085124}
}

@article{Lieb1961,
  title={Two soluble models of an antiferromagnetic chain},
  author={Lieb, Elliott and Schultz, Theodore and Mattis, Daniel},
  journal={Annals of Physics},
  volume={16},
  number={3},
  pages={407--466},
  year={1961},
  publisher={Elsevier}
}

@article{Haldane2004,
  author  = {Haldane, F. D. M.},
  title   = {{Berry} Curvature on the {Fermi} Surface:
             Anomalous {Hall} Effect as a Topological
             {Fermi}-Liquid Property},
  journal = {Phys. Rev. Lett.},
  volume  = {93},
  number  = {20},
  pages   = {206602},
  year    = {2004},
  doi     = {10.1103/PhysRevLett.93.206602}
}

@article{Inbar2023,
  author  = {Inbar, Alon and Birkbeck, John and Xiao, Jiewen
             and Taniguchi, Takashi and Watanabe, Kenji
             and Yan, Binghai and Oreg, Yuval and Stern, Ady
             and Berg, Erez and Ilani, Shahal},
  title   = {The quantum twisting microscope},
  journal = {Nature},
  volume  = {614},
  number  = {7949},
  pages   = {682--687},
  year    = {2023},
  doi     = {10.1038/s41586-022-05685-y}
}

@incollection{Truesdell1960,
  author    = {Truesdell, Clifford A. and Toupin, Richard A.},
  title     = {The Classical Field Theories},
  booktitle = {Principles of Classical Mechanics and Field Theory (Prinzipien der Klassischen Mechanik und Feldtheorie)},
  editor    = {Fl{\"{u}}gge, Siegfried},
  series    = {Handbuch der Physik},
  volume    = {III/1},
  pages     = {},
  year      = {1960},
  publisher = {Springer-Verlag},
  address   = {Berlin},
  doi       = {10.1007/978-3-642-45943-6_2}
}

@book{Slattery1990,
  title     = {Interfacial Transport Phenomena},
  author    = {Slattery, John Charles},
  year      = {1990},
  publisher = {Springer-Verlag},
  address   = {New York, NY},
  isbn      = {978-0-387-97387-6}
}

@book{Mahan2000,
  author    = {Mahan, Gerald D.},
  title     = {Many-Particle Physics},
  edition   = {3rd},
  publisher = {Kluwer Academic/Plenum},
  address   = {New York},
  year      = {2000}
}

@book{PinesNozieres1966,
  author    = {Pines, David and Nozi{\`e}res, Philippe},
  title     = {The Theory of Quantum Liquids, Vol.~I: Normal {F}ermi Liquids},
  publisher = {W.~A.~Benjamin},
  address   = {New York},
  year      = {1966}
}

@book{Aris1962,
  author    = {Aris, Rutherford},
  title     = {Vectors, Tensors, and the Basic Equations of Fluid Mechanics},
  publisher = {Prentice-Hall},
  address   = {Englewood Cliffs, NJ},
  year      = {1962}
}

@article{Luttinger1961,
  author  = {Luttinger, J. M.},
  title   = {Analytic Properties of Single-Particle Propagators for Many-Fermion Systems},
  journal = {Phys. Rev.},
  volume  = {121},
  pages   = {942--949},
  year    = {1961},
  doi     = {10.1103/PhysRev.121.942},
  url     = {https://doi.org/10.1103/PhysRev.121.942}
}

@article{Migdal1957,
  author  = {Migdal, A. B.},
  title   = {The momentum distribution of interacting {F}ermi particles},
  journal = {Sov. Phys. JETP},
  volume  = {5},
  pages   = {333},
  year    = {1957},
  note    = {[Zh. Eksp. Teor. Fiz. 32, 399 (1957)]}
}

@article{BaymKadanoff1961,
  author  = {Baym, Gordon and Kadanoff, Leo P.},
  title   = {Conservation Laws and Correlation Functions},
  journal = {Phys. Rev.},
  volume  = {124},
  pages   = {287--299},
  year    = {1961},
  doi     = {10.1103/PhysRev.124.287},
  url     = {https://doi.org/10.1103/PhysRev.124.287}
}

@article{Baym1962,
  author  = {Baym, Gordon},
  title   = {Self-Consistent Approximations in Many-Body Systems},
  journal = {Phys. Rev.},
  volume  = {127},
  pages   = {1391--1401},
  year    = {1962},
  doi     = {10.1103/PhysRev.127.1391},
  url     = {https://doi.org/10.1103/PhysRev.127.1391}
}

@article{KozikFerreroGeorges2015,
  author  = {Kozik, Evgeny and Ferrero, Michel and Georges, Antoine},
  title   = {Nonexistence of the {L}uttinger-{W}ard functional and misleading convergence of skeleton diagrammatic series for {H}ubbard-like models},
  journal = {Phys. Rev. Lett.},
  volume  = {114},
  pages   = {156402},
  year    = {2015},
  doi     = {10.1103/PhysRevLett.114.156402}
}

\clearpage
\onecolumngrid
\setcounter{equation}{0}
\setcounter{section}{0}
\setcounter{subsection}{0}
\setcounter{secnumdepth}{2}
\renewcommand{\theequation}{S\arabic{equation}}
\renewcommand{\thesection}{S\arabic{section}}
\renewcommand{\thesubsection}{\thesection.\arabic{subsection}}
\renewcommand{\theHequation}{S\arabic{equation}}
\renewcommand{\theHsection}{S\arabic{section}}
\makeatletter\renewcommand{\p@subsection}{}\makeatother
\allowdisplaybreaks
\begin{center}
{\large\bfseries Supplemental Material for ``Exact solutions for kinematics across interaction-deformed Fermi surfaces''}\\[1.2ex]
Alexander Kruchkov and F. Duncan M. Haldane\\[0.4ex]
{\itshape Department of Physics, Princeton University, Princeton NJ 08544, USA}
\end{center}
\vspace{1ex}

Luttinger's theorem~\cite{Luttinger1960,LuttingerWard1960} states that the volume enclosed by the Fermi surface of an interacting Fermi liquid equals the particle density, however much the interaction reshapes the surface. This Supplemental Material derives the identities of the main text, which describe how the momentum distribution rearranges itself around the moving Fermi surface as the interaction is switched on or modified. Sections~\ref{sec:S_spectral}--\ref{sec:S_count} define the Green's function, the interacting Fermi surface, Migdal's discontinuity, Luttinger's integral, and the quasihole deficit and quasiparticle surplus. Sections~\ref{sec:S_geometry}--\ref{sec:S_balance} derive the kinematic identities, which hold whether or not Luttinger's theorem does. Section~\ref{sec:S_LW} adds Luttinger's theorem and derives its consequences.

\section{Interacting Green's functions, spectral functions, and momentum distributions}
\label{sec:S_spectral}

We consider spinless fermions in a single band on a $d$-dimensional lattice of $\mathcal{N}$ sites. The operator $c^{\dagger}_{\vec k}$ creates a fermion in the Bloch state of crystal momentum $\vec k$ in the Brillouin zone (BZ), and $\omega_{\vec k}$ is the band energy. The interaction is switched on along a family of Hamiltonians
\begin{equation}
 H_{\lambda} = \sum_{\vec k} \omega_{\vec k}\, c^{\dagger}_{\vec k} c_{\vec k} + V_{\lambda} ,
 \qquad 0 \le \lambda \le 1 ,
 \label{eq:S_hamiltonian}
\end{equation}
in which $V_{\lambda}$ is a translation-invariant interaction that conserves the particle number and depends smoothly on the coupling $\lambda$, with $V_{0} = 0$ (free fermions) and $V_{1} = V$ (the physical system).   At each $\lambda$ the system is in the ground state $|0,N\rangle$ of $H_{\lambda}$ with $N$ particles and energy $E^{N}_{0}$ (the dependence on $\lambda$ is not written). The density $N/\mathcal{N}$ is held fixed, so that the chemical potential, $\mu = E^{N+1}_{0} - E^{N}_{0}$ (equal to $E^{N}_{0} - E^{N-1}_{0}$ in the thermodynamic limit of a metal), changes with the coupling, $\mu = \mu(\lambda)$. We call this fixed-density family of ground states the coupling path, or simply $\lambda$-path or the path.    We set $\hbar = k_{B} = 1$ and the volume of the unit cell to one, and work at zero temperature in the thermodynamic limit of an interacting system, where momentum sums become Brillouin-zone integrals,
$\mathcal{N}^{-1}\sum_{\vec k} \to \intk$, normalized so that $\intk 1 = 1$; all particle numbers below are numbers per site. Frequencies are measured from the fixed zero of energy of $H_{\lambda}$, not from the chemical potential, which enters only through the occupation of the states.

The retarded Green's function of a fermion with momentum $\vec k$ is~\cite{AGD1963,Mahan2000}
\begin{equation}
 \G^{R}_{\lambda}(\omega,\vec k) = -i\int_{0}^{\infty} dt\, e^{i(\omega + i0^{+})t}\,
 \langle 0,N|\bigl\{c_{\vec k}(t), c^{\dagger}_{\vec k}\bigr\}|0,N\rangle ,
 \qquad
 c_{\vec k}(t) = e^{iH_{\lambda}t}\, c_{\vec k}\, e^{-iH_{\lambda}t} ,
 \label{eq:S_GRdef}
\end{equation}
where $\{\,,\}$ is the anticommutator. To evaluate it we insert the eigenstates $|m,N{\pm}1\rangle$ of $H_{\lambda}$ with $N \pm 1$ particles and energies $E^{N\pm1}_{m}$, and introduce the addition energies $\epsilon^{+}_{m} = E^{N+1}_{m} - E^{N}_{0}$ and the removal energies $\epsilon^{-}_{m} = E^{N}_{0} - E^{N-1}_{m}$, the energy carried off by the removed particle. The time integral then gives $\G^{R}_{\lambda}(\omega,\vec k) = \int d\epsilon\, A_{\lambda}(\epsilon,\vec k)/(\omega - \epsilon + i0^{+})$, with the spectral function in the Lehmann representation
\begin{equation}
 A_{\lambda}(\omega,\vec k) = \sum_{m} \bigl|\langle m,N{+}1|c^{\dagger}_{\vec k}|0,N\rangle\bigr|^{2}\,\delta(\omega - \epsilon^{+}_{m})
 + \sum_{m} \bigl|\langle m,N{-}1|c_{\vec k}|0,N\rangle\bigr|^{2}\,\delta(\omega - \epsilon^{-}_{m}) ,
 \label{eq:S_lehmann}
\end{equation}
where the first sum describes the addition of a particle and the second its removal. Both have non-negative weights, and completeness together with $\{c_{\vec k}, c^{\dagger}_{\vec k}\} = 1$ gives the sum rule
\begin{equation}
 \int_{-\infty}^{\infty} d\epsilon\, A_{\lambda}(\epsilon,\vec k) = 1 .
 \label{eq:S_sumrule}
\end{equation}
A similar integral form defines the \textit{interacting}  Green's function at any complex frequency $z$ off the real axis,
\begin{equation}
 \G_{\lambda}(z,\vec k) = \int_{-\infty}^{\infty} d\epsilon\, \frac{A_{\lambda}(\epsilon,\vec k)}{z - \epsilon} ,
 \qquad
 A_{\lambda}(\epsilon,\vec k) = -\frac{1}{\pi}\Im\G^{R}_{\lambda}(\epsilon,\vec k) \ge 0 .
 \label{eq:S_spectral}
\end{equation}

 The interacting occupation number  is 
\begin{equation}
 n_{\lambda\vec k} = \int_{-\infty}^{\mu} d\omega\, A_{\lambda}(\omega,\vec k)
 = -\frac{1}{\pi}\Im\int_{-\infty}^{\mu} d\omega\, \G^{R}_{\lambda}(\omega,\vec k) .
 \label{eq:S_nzero}
\end{equation}
 Positivity of $A_{\lambda}$ and the sum rule~\eqref{eq:S_sumrule} give
\begin{equation}
 0 \le n_{\lambda\vec k} \le 1 .
 \label{eq:S_nbounds}
\end{equation}

The main text also introduces  the Matsubara (finite temperature) form of the momentum distribution. At temperature $T = 1/\beta$ the imaginary-time Green's function is $\G_{\lambda}(\tau,\vec k) = -\langle T_{\tau}\, c_{\vec k}(\tau)\, c^{\dagger}_{\vec k}\rangle$, where $c_{\vec k}(\tau) = e^{\tau K_{\lambda}} c_{\vec k}\, e^{-\tau K_{\lambda}}$ with $K_{\lambda} = H_{\lambda} - \mu\hat N$ and $\hat N = \sum_{\vec k} c^{\dagger}_{\vec k}c_{\vec k}$, $T_{\tau}$ orders the operators in imaginary time, and $\langle\cdots\rangle = \mathrm{Tr}(e^{-\beta K_{\lambda}}\cdots)/\mathrm{Tr}\,e^{-\beta K_{\lambda}}$~\cite{Mahan2000}. Its value at $\tau = 0^{-}$ is $\langle c^{\dagger}_{\vec k} c_{\vec k}\rangle = n_{\lambda\vec k}$, and its Fourier coefficients at the fermionic Matsubara frequencies $\omega_{n}$ are $\G_{\lambda}(\mu + i\omega_{n},\vec k)$, the shift by $\mu$ arising because our frequencies are measured from the fixed zero of energy. Therefore [Eq.~\eqref{eq:setup_matsubara} of the main text]
\begin{equation}
 n_{\lambda\vec k} = \frac{1}{\beta}\sum_{\omega_{n}} \G_{\lambda}(\mu + i\omega_{n},\vec k)\,e^{i\omega_{n}0^{+}} ,
 \qquad
 \omega_{n} = (2n + 1)\pi/\beta .
 \label{eq:S_matsubara}
\end{equation}
The frequency sum is done with the elementary result, valid for $0 < \tau < \beta$ and a real energy $\xi$, measured from the chemical potential ($\xi = \epsilon - \mu$ below),
\begin{equation}
 \frac{1}{\beta}\sum_{\omega_{n}} \frac{e^{i\omega_{n}\tau}}{i\omega_{n} - \xi} = e^{\xi\tau} f_{\beta}(\xi) ,
 \qquad
 f_{\beta}(\xi) = \frac{1}{e^{\beta\xi} + 1} .
 \label{eq:S_sum}
\end{equation}
To prove it, consider the function $g_{\xi}(s) = -[1 - f_{\beta}(\xi)]\,e^{-\xi s}$ for $0 < s < \beta$, continued to other values of $s$ antiperiodically, $g_{\xi}(s - \beta) = -g_{\xi}(s)$. Its Fourier coefficients are
\begin{equation}
 \int_{0}^{\beta} ds\, e^{i\omega_{n}s} g_{\xi}(s)
 = \frac{[1 - f_{\beta}(\xi)](1 + e^{-\beta\xi})}{i\omega_{n} - \xi}
 = \frac{1}{i\omega_{n} - \xi} ,
 \label{eq:S_sumproof}
\end{equation}
since $e^{i\omega_{n}\beta} = -1$ and $[1 - f_{\beta}(\xi)](1 + e^{-\beta\xi}) = 1$. The Fourier series then gives $\beta^{-1}\sum_{n} e^{i\omega_{n}\tau}/(i\omega_{n} - \xi) = g_{\xi}(-\tau) = -g_{\xi}(\beta - \tau) = e^{\xi\tau}f_{\beta}(\xi)$, which is Eq.~\eqref{eq:S_sum}. At $T > 0$ the Green's function and the spectral function are the thermal ones, with the same spectral representation~\eqref{eq:S_spectral}~\cite{Mahan2000}. Inserting it into Eq.~\eqref{eq:S_matsubara}, with $\xi = \epsilon - \mu$, and letting $\tau \to 0^{+}$ gives
\begin{equation}
 n_{\lambda\vec k} = \int_{-\infty}^{\infty} d\epsilon\, A_{\lambda}(\epsilon,\vec k)\, f_{\beta}(\epsilon - \mu) .
 \label{eq:S_nfiniteT}
\end{equation}
As $T \to 0$ the thermal quantities become those of the ground state, Eqs.~\eqref{eq:S_lehmann} and \eqref{eq:S_nzero}, and Eq.~\eqref{eq:S_nfiniteT} reduces to Eq.~\eqref{eq:S_nzero}; only this limit is used below.

The self-energy $\Sigma_{\lambda}$ is defined by Dyson's equation,
\begin{equation}
 \G_{\lambda}(z,\vec k)^{-1} = z - \omega_{\vec k} - \Sigma_{\lambda}(z,\vec k) ,
 \qquad
 \G^{R}_{\lambda}(\omega,\vec k) = \G_{\lambda}(\omega + i0^{+},\vec k) .
 \label{eq:S_dyson}
\end{equation}
At $\lambda = 0$ the self-energy vanishes, $\Sigma_{0} = 0$, and $\Sigma^{R}_{\lambda}(\omega,\vec k) = \Sigma_{\lambda}(\omega + i0^{+},\vec k)$ is the retarded self-energy. Since $\G_{\lambda}$ has no zeros off the real axis, $\Sigma_{\lambda} = z - \omega_{\vec k} - \G^{-1}_{\lambda}$ is analytic there; for $\eta > 0$, $\Im\Sigma_{\lambda}(\omega + i\eta,\vec k) = \eta - \Im\G^{-1}_{\lambda}(\omega + i\eta,\vec k) < \eta$, and letting $\eta \to 0^{+}$ gives $\Im\Sigma^{R}_{\lambda}(\omega,\vec k) \le 0$~\cite{AGD1963, Luttinger1961}. Its behavior at large frequency follows from the moments $M_{j} = \int d\epsilon\,\epsilon^{j}A_{\lambda}(\epsilon,\vec k)$ of the spectral function, which we assume to be finite, uniformly in $\vec k$ and in the size of the system (the expansion below uses them up to $j = 3$). It holds for a bounded interaction of finite range, such as a nearest-neighbor density--density interaction, whereas for the Coulomb interaction already the Hartree part of $M_{1}$ is finite only with a neutralizing background. With finite moments, expanding Eq.~\eqref{eq:S_spectral} in powers of $1/z$ and inverting gives
\begin{equation}
 \G_{\lambda}(z,\vec k) = \frac{1}{z} + \frac{M_{1}}{z^{2}} + \frac{M_{2}}{z^{3}} + \dots ,
 \qquad
 \Sigma_{\lambda}(z,\vec k) = \Sigma^{\infty}_{\lambda\vec k} + \frac{M_{2} - M_{1}^{2}}{z} + O(z^{-2}) ,
 \label{eq:S_highfreq}
\end{equation}
where $\Sigma^{\infty}_{\lambda\vec k} = M_{1} - \omega_{\vec k}$ and $M_{1} = \langle 0,N|\{[c_{\vec k},H_{\lambda}], c^{\dagger}_{\vec k}\}|0,N\rangle$. For a two-body interaction $\Sigma^{\infty}$ is the instantaneous (Hartree--Fock) self-energy evaluated with the exact occupations. Thus $\G_{\lambda} \simeq 1/z$, $\Sigma_{\lambda}$ stays bounded, and $\partial_{z}\Sigma_{\lambda} = O(z^{-2})$ at large $|z|$: the frequency integrals below converge, and the boundary terms of $\G_{\lambda}\Sigma_{\lambda}$ and $\G_{\lambda}\,\partial_{z}\Sigma_{\lambda}$ at $|\omega| \to \infty$ vanish. The coupling derivatives need one more condition, which we assume: Eq.~\eqref{eq:S_highfreq} holds uniformly along the path and can be differentiated with respect to $\lambda$, so that $\partial_{\lambda}\Sigma_{\lambda}(z,\vec k) = \partial_{\lambda}\Sigma^{\infty}_{\lambda\vec k} + O(1/z)$ stays bounded and
\begin{equation}
 \G_{\lambda}(z,\vec k)\,\partial_{\lambda}\Sigma_{\lambda}(z,\vec k) \longrightarrow 0 ,
 \qquad |z| \to \infty ;
 \label{eq:S_uvcondition}
\end{equation}
and the frequency integrals for $n_{\lambda\vec k}$ and $\LL_{\lambda\vec k}$ below can be differentiated with respect to $\lambda$ under the integral sign. Equation~\eqref{eq:S_uvcondition} is the lower-endpoint condition used below Eq.~\eqref{eq:S_Lderivintegrand}.

\section{Fermi surface, quasiparticle pole, and Migdal's discontinuity}
\label{sec:S_pole}

In a Fermi liquid the quasiparticles at the Fermi level are undamped. We assume throughout that the self-energy is real at the chemical potential, $\Im\Sigma^{R}_{\lambda}(\mu,\vec k) = 0$, and that $\Sigma^{R}_{\lambda}(\omega,\vec k)$ is differentiable in $\omega$ near the Fermi level and in $\lambda$ along the path. The inverse Green's function at the chemical potential is then a real function of momentum [Eq.~\eqref{eq:flow_phi} of the main text],
\begin{equation}
 \Phi_{\lambda\vec k} = \G^{-1}_{\lambda}(\mu + i0^{+},\vec k) = \mu - \omega_{\vec k} - \Re\Sigma^{R}_{\lambda}(\mu,\vec k) ,
 \label{eq:S_phi}
\end{equation}
which plays the role of $\mu - \omega_{\vec k}$ for free fermions~\cite{Luttinger1960,Dzyaloshinskii2003}. The main text writes this real value of the self-energy as $\Sigma_{\lambda}(\mu,\vec k)$ or $\Sigma^{R}_{\lambda}(\mu,\vec k)$, and $\partial_{\omega}\Sigma_{\lambda}(\mu,\vec k)$ and $\partial_{\lambda}\Sigma_{\lambda}(\mu,\vec k)$ for the derivatives of $\Sigma^{R}_{\lambda}$ at the Fermi level, which are also real (Sec.~\ref{sec:S_rates}). The Fermi sea $\Omega_{\lambda}$ is the region where it is positive, and the Fermi surface $\FS_{\lambda}(\vec k)$ is where it vanishes,
\begin{equation}
 \Omega_{\lambda} (\vec k)  = \{\vec k : \Phi_{\lambda\vec k} > 0\} ,
 \qquad
 \FS_{\lambda} (\vec k) = \{\vec k : \Phi_{\lambda\vec k} = 0\} ;
 \label{eq:S_domain}
\end{equation}
we denote by $\Omega^{c}_{\lambda}$ the rest of the Brillouin zone, outside the Fermi surface, where $\Phi_{\lambda\vec k} < 0$. With $\Theta$ the Heaviside step function, $\Theta(\Phi_{\lambda\vec k})$ is the occupation of a sharp Fermi sea bounded by the interacting Fermi surface; for free fermions it is the familiar step $\Theta(\mu - \omega_{\vec k})$. We assume that $\FS_{\lambda}(\vec k)$ is a smooth surface, possibly with several sheets, on which $\nabla_{\vec k}\Phi_{\lambda\vec k} \ne 0$; that its topology does not change along the $\lambda$-path; and that $\G_{\lambda}(\mu,\vec k)$ has no zeros, so that $\Phi_{\lambda\vec k}$ is finite and changes sign only on $\FS_{\lambda}$. The surface has zero volume and does not contribute to Brillouin-zone integrals. For a quantity $F_{\lambda\vec k}$ that jumps at the Fermi surface and a point $\vec k \in \FS_{\lambda}$, $F^{-}_{\lambda\vec k}$ and $F^{+}_{\lambda\vec k}$ denote the limits of $F_{\lambda\vec k'}$ as $\vec k' \to \vec k$ from inside and from outside the sea, and the jump is
\begin{equation}
 [F_{\lambda\vec k}] = F^{-}_{\lambda\vec k} - F^{+}_{\lambda\vec k} .
 \label{eq:S_jumpdef}
\end{equation}

Near the Fermi surface the Green's function is dominated by the quasiparticle pole~\cite{Nozieres1964,Mahan2000}. We define the real on-shell quasiparticle energy $\tilde\omega_{\lambda\vec k}$, the real root near $\mu$ of $\omega - \omega_{\vec k} - \Re\Sigma^{R}_{\lambda}(\omega,\vec k) = 0$, and the quasiparticle weight $Z_{\lambda\vec k}$ by [Eq.~\eqref{eq:setup_Z} of the main text]:

\begin{equation}
 \tilde\omega_{\lambda\vec k} = \omega_{\vec k} + \Re\Sigma^{R}_{\lambda}(\tilde\omega_{\lambda\vec k},\vec k) ,
 \qquad
 Z^{-1}_{\lambda\vec k} = 1 - \partial_{\omega}\Re\Sigma^{R}_{\lambda}(\tilde\omega_{\lambda\vec k},\vec k) .
 \label{eq:S_qp}
\end{equation}
On the Fermi surface $\tilde\omega_{\lambda\vec k} = \mu$, and $Z_{\lambda\vec k}$ is the weight of the undamped quasiparticle singularity of $\G^{R}_{\lambda}$ at $\omega = \mu$, $\lim_{\omega\to\mu}(\omega - \mu)\,\G^{R}_{\lambda}(\omega,\vec k) = Z_{\lambda\vec k}$, because the self-energy and its frequency derivative are real there: $\Im\Sigma^{R}_{\lambda} \le 0$ has its maximum, zero, at $\omega = \mu$. Near the Fermi surface but off it, $Z_{\lambda\vec k}$ is an on-shell weight, and Eq.~\eqref{eq:S_pole} below is the leading small-damping approximation. If $\Sigma^{R}_{\lambda}$ has an analytic continuation $\Sigma^{\mathrm{cont}}_{\lambda}$ below the real axis and the continued propagator $[z - \omega_{\vec k} - \Sigma^{\mathrm{cont}}_{\lambda}(z,\vec k)]^{-1}$ has a simple pole $z_{p}$ there, $z_{p} - \omega_{\vec k} - \Sigma^{\mathrm{cont}}_{\lambda}(z_{p},\vec k) = 0$, its exact residue $Z_{p} = [1 - \partial_{z}\Sigma^{\mathrm{cont}}_{\lambda}(z_{p},\vec k)]^{-1}$ is in general complex and need not equal $Z_{\lambda\vec k}$; such a continuation does not always exist. Like $n_{\lambda\vec k}$ and $\Phi_{\lambda\vec k}$, $Z_{\lambda\vec k}$ depends on $\lambda$. Expanding $\G^{R}_{\lambda}(\omega,\vec k)^{-1}$ to first order in $\omega - \tilde\omega_{\lambda\vec k}$ gives the quasiparticle form

\begin{equation}
 \G^{R}_{\lambda}(\omega,\vec k) \simeq \frac{Z_{\lambda\vec k}}{\omega - \tilde\omega_{\lambda\vec k} + i\Gamma_{\lambda\vec k}} ,
 \qquad
 \frac{\Gamma_{\lambda\vec k}}{|\tilde\omega_{\lambda\vec k} - \mu|} \longrightarrow 0
 \quad \text{as } \vec k \to \FS_{\lambda} ,
 \label{eq:S_pole}
\end{equation}
with the damping $\Gamma_{\lambda\vec k} = -Z_{\lambda\vec k}\Im\Sigma^{R}_{\lambda}(\tilde\omega_{\lambda\vec k},\vec k) \ge 0$. The condition on the right is the Fermi-liquid property we assume: the damping vanishes faster than the excitation energy.   Equation~\eqref{eq:S_pole} describes only the singular part of the Green's function. The full $\G^{R}_{\lambda}$ also has an incoherent part, the spectral weight outside the quasiparticle peak, and it is the full spectral function that enters Eq.~\eqref{eq:S_nzero}.

 We further assume that the incoherent part of the spectral function varies continuously across the Fermi surface, so that its occupied weight has the same limit $n_{\mathrm{inc},\FS}$ on both sides. The occupations just inside and just outside the Fermi surface are then

\begin{equation}
 n^{-}_{\lambda\vec k} = Z_{\lambda\vec k} + n_{\mathrm{inc},\FS} ,
 \qquad
 n^{+}_{\lambda\vec k} = n_{\mathrm{inc},\FS} ,
 \qquad
 [n_{\lambda\vec k}] = n^{-}_{\lambda\vec k} - n^{+}_{\lambda\vec k} = Z_{\lambda\vec k} ,
 \label{eq:S_migdal}
\end{equation}
which is Migdal's discontinuity~\cite{Migdal1957,Luttinger1960}: the momentum distribution drops by the quasiparticle residue across the interacting Fermi surface. The residue of a quasiparticle pole is positive and, by Eq.~\eqref{eq:S_nbounds}, at most one, $0 < Z_{\lambda\vec k} \le 1$ on $\FS_{\lambda}$; the hole occupation just inside the surface is

\begin{equation}
 1 - n^{-}_{\lambda\vec k} = 1 - Z_{\lambda\vec k} - n^{+}_{\lambda\vec k} \ge 0 .
 \label{eq:S_traceweights}
\end{equation}
For free fermions $Z_{\lambda\vec k} = 1$, $n^{-}_{\lambda\vec k} = 1$, and $n^{+}_{\lambda\vec k} = 0$.

The function $\Phi_{\lambda\vec k}$ is not the quasiparticle energy. It equals $\mu - \varepsilon^{*}_{\lambda\vec k}$, where $\varepsilon^{*}_{\lambda\vec k} = \omega_{\vec k} + \Re\Sigma^{R}_{\lambda}(\mu,\vec k)$ is the band shifted by the self-energy at the Fermi level, which coincides with $\tilde\omega_{\lambda\vec k}$ only on $\FS_{\lambda}$. Expanding the self-energy in Eq.~\eqref{eq:S_qp} about $\omega = \mu$ gives, near the surface,

\begin{equation}
 \Phi_{\lambda\vec k} = \frac{\mu - \tilde\omega_{\lambda\vec k}}{Z_{\lambda\vec k}} + o\bigl(|\mu - \tilde\omega_{\lambda\vec k}|\bigr) ,
 \label{eq:S_philinear}
\end{equation}
so that $\Phi > 0$ where the quasiparticle level lies below the chemical potential. Differentiating Eq.~\eqref{eq:S_qp} with respect to momentum, with the frequency argument of the self-energy held fixed in its gradient, gives the quasiparticle velocity,

\begin{equation}
 \nabla_{\vec k}\tilde\omega_{\lambda\vec k} = Z_{\lambda\vec k}\bigl[\nabla_{\vec k}\omega_{\vec k} + \partial_{\vec k}\Re\Sigma^{R}_{\lambda}(\tilde\omega_{\lambda\vec k},\vec k)\bigr] ,
 \label{eq:S_qpgradient}
\end{equation}
and on the Fermi surface, where $\tilde\omega_{\lambda\vec k} = \mu$, comparison with the gradient of Eq.~\eqref{eq:S_phi} gives the exact relations

\begin{equation}
 \nabla_{\vec k}\Phi_{\lambda\vec k} = -\frac{\nabla_{\vec k}\tilde\omega_{\lambda\vec k}}{Z_{\lambda\vec k}} ,
 \qquad
 \frac{dS_{\vec k}}{|\nabla_{\vec k}\Phi_{\lambda\vec k}|} = \frac{Z_{\lambda\vec k}\, dS_{\vec k}}{|\nabla_{\vec k}\tilde\omega_{\lambda\vec k}|} .
 \label{eq:S_gradient}
\end{equation}
Here $dS_{\vec k}$ is the element of area of the Fermi surface, and $dS_{\vec k}/[(2\pi)^{d}|\nabla_{\vec k}\tilde\omega_{\lambda\vec k}|]$ is the weight of the quasiparticle density of states on it (Sec.~\ref{sec:S_mudrift}).

\section{Luttinger's integral and the local form of the momentum distribution}
\label{sec:S_local}

Luttinger's integrand is the momentum-resolved quantity  ~\cite{Luttinger1960,LuttingerWard1960} [Eq.~\eqref{eq:antiflow_IL} of the main text]
\begin{equation}
 \LL_{\lambda\vec k} = \frac{1}{\pi}\Im\int_{-\infty}^{\mu} d\omega\, \G^{R}_{\lambda}(\omega,\vec k)\,\partial_{\omega}\Sigma^{R}_{\lambda}(\omega,\vec k) .
 \label{eq:S_Ldef}
\end{equation}
Here $\G^{R}_{\lambda}$ and $\Sigma^{R}_{\lambda}$ are taken at the same complex frequency $\omega + i\eta$, $\eta \to 0^{+}$. Its Brillouin-zone integral is the term that Luttinger and Ward showed to vanish (Sec.~\ref{sec:S_LW}); here we need its value at each momentum. The full Green's function must be kept in Eq.~\eqref{eq:S_Ldef}: the quasiparticle pole alone does not determine $\LL_{\lambda\vec k}$. The link between $\LL$ and $n$ is the identity that underlies the proofs of Luttinger's theorem~\cite{Luttinger1960,SekiYunoki2017}. Dyson's equation~\eqref{eq:S_dyson} gives, at any complex frequency,
\begin{align}
 \partial_{z}\ln\G^{-1}_{\lambda} & = \G_{\lambda}\,\partial_{z}\G^{-1}_{\lambda} = \G_{\lambda}(1 - \partial_{z}\Sigma_{\lambda}) ,
\\
 \G_{\lambda} & = \partial_{z}\ln\G^{-1}_{\lambda} + \G_{\lambda}\,\partial_{z}\Sigma_{\lambda} .
 \label{eq:S_logidentity}
\end{align}
The first term on the right hand side of Eq \eqref{eq:S_logidentity} is a total derivative, and its integral is fixed by the phase of $\G^{-1}_{\lambda}$. This phase is unambiguous: by Eq.~\eqref{eq:S_spectral}, $\Im\G^{-1}_{\lambda} > 0$ throughout the upper half-plane, so the logarithm of $\G^{-1}_{\lambda}$ is analytic there and its phase $\arg\G^{-1}_{\lambda}$ lies between $0$ and $\pi$. (A fixed energy scale that makes the argument of the logarithm dimensionless drops out of the differences below.)

We now integrate Eq.~\eqref{eq:S_logidentity} along the line $z = \omega + i\eta$ from $\omega = -R$ to $\omega = \mu$, take $-\pi^{-1}\Im$ of the result, and let $R \to \infty$ and then $\eta \to 0^{+}$. By Eqs.~\eqref{eq:S_nzero} and \eqref{eq:S_Ldef} the left side becomes $n_{\lambda\vec k}$ and the last term $-\LL_{\lambda\vec k}$:

\begin{align}
 n_{\lambda\vec k} &= Q_{\lambda\vec k} - \LL_{\lambda\vec k} ,
 \label{eq:S_nQL} \\
 Q_{\lambda\vec k} &= -\frac{1}{\pi}\lim_{\eta\to0^{+}}\lim_{R\to\infty}
 \Im\bigl[\ln\G^{-1}_{\lambda}(\mu + i\eta,\vec k) - \ln\G^{-1}_{\lambda}(-R + i\eta,\vec k)\bigr] .
 \label{eq:S_Qdef}
\end{align}
The imaginary part is taken before the limits: the real part of the bracket, $\ln|\G^{-1}_{\lambda}(\mu + i\eta,\vec k)| - \ln|\G^{-1}_{\lambda}(-R + i\eta,\vec k)|$, diverges as $-\ln R$, so the difference of the logarithms itself has no limit, whereas its imaginary part, the change of the phase of $\G^{-1}_{\lambda}$, has a finite one.
At the lower end $\G^{-1}_{\lambda}(z,\vec k) \simeq z$ by Eq.~\eqref{eq:S_highfreq}, so the phase is $\pi$; this fixes the branch of the logarithm. At the upper end, off the Fermi surface, $\G^{-1}_{\lambda}(\mu + i0^{+},\vec k) = \Phi_{\lambda\vec k}$ is real and finite, with phase $0$ in the sea and $\pi$ outside. Hence

\begin{equation}
 Q_{\lambda\vec k} = 1 - \frac{1}{\pi}\arg\G^{-1}_{\lambda}(\mu + i0^{+},\vec k)
 = \begin{cases} 1 , & \Phi_{\lambda\vec k} > 0 , \\ 0 , & \Phi_{\lambda\vec k} < 0 . \end{cases}
 \label{eq:S_Qstep}
\end{equation}
The momentum distribution away from the Fermi surface is therefore [Eq.~\eqref{eq:antiflow_pointwise} of the main text]
\begin{equation}
 n_{\lambda\vec k} = \Theta(\Phi_{\lambda\vec k}) - \LL_{\lambda\vec k} ,
 \qquad \vec k \notin \FS_{\lambda} .
 \label{eq:S_local}
\end{equation}
The occupation of the interacting system is the sharp step at its Fermi surface minus Luttinger's integral. It uses only causality, the large-frequency behavior~\eqref{eq:S_highfreq}, and the reality of $\Sigma^{R}_{\lambda}(\mu,\vec k)$; nothing is assumed about the zone integral of $\LL$. For free fermions $\LL = 0$, and Eq.~\eqref{eq:S_local} is the Fermi step. Since $0 \le n \le 1$, the decomposition fixes the sign of $\LL$ on either side of the surface,
\begin{equation}
 \LL_{\lambda\vec k} = \begin{cases} 1 - n_{\lambda\vec k} \ge 0 , & \vec k \in \Omega_{\lambda} , \\ -n_{\lambda\vec k} \le 0 , & \vec k \in \Omega^{c}_{\lambda} . \end{cases}
 \label{eq:S_Lsign}
\end{equation}
Luttinger's integral is the hole occupation $1 - n$ inside the sea and minus the particle occupation outside it. At the Fermi surface, Eqs.~\eqref{eq:S_migdal} and \eqref{eq:S_Lsign} give two different jumps,

\begin{align}
 [\LL_{\lambda\vec k}] &= (1 - n^{-}_{\lambda\vec k}) - (-n^{+}_{\lambda\vec k}) = 1 - Z_{\lambda\vec k} ,
 \label{eq:S_Ljump} \\
 [|\LL_{\lambda\vec k}|] &= (1 - n^{-}_{\lambda\vec k}) - n^{+}_{\lambda\vec k} = 1 - Z_{\lambda\vec k} - 2n^{+}_{\lambda\vec k} .
 \label{eq:S_absjump}
\end{align}
The jump of $\LL$ is the incoherent weight $1 - Z_{\lambda\vec k} \ge 0$; the jump of $|\LL|$ is the difference of two non-negative limits and has no definite sign.

\section{Quasihole deficit and quasiparticle surplus}
\label{sec:S_count}

The Fermi volume $V_{F}$, the fraction of the zone occupied by the Fermi sea, and the density $N_{\mathrm{tot}}$ are [Eq.~\eqref{eq:setup_VF} of the main text]
\begin{equation}
 V_{F}(\lambda) = \intk \Theta(\Phi_{\lambda\vec k}) ,
 \qquad
 N_{\mathrm{tot}}(\lambda) = \intk n_{\lambda\vec k} .
 \label{eq:S_VN}
\end{equation}
Luttinger's theorem states that $V_{F} (\lambda) = N_{\mathrm{tot}} (\lambda)$~\cite{Luttinger1960,LuttingerWard1960}. Its content is best expressed through the occupation missing inside the sea and the occupation present outside it [Eq.~\eqref{eq:setup_N1} of the main text],
\begin{align}
 N_{1}(\lambda) &= \intin (1 - n_{\lambda\vec k}) = \intk \Theta(\Phi_{\lambda\vec k})\,(1 - n_{\lambda\vec k}) ,
 \label{eq:S_N1} \\
 N_{2}(\lambda) &= \intout n_{\lambda\vec k} = \intk \Theta(-\Phi_{\lambda\vec k})\, n_{\lambda\vec k} .
 \label{eq:S_N2}
\end{align}
We call $N_{1}$ the quasihole deficit and $N_{2}$ the quasiparticle surplus. Both are non-negative properties of the ground-state momentum distribution: they measure its deviation from the step $\Theta(\Phi_{\lambda\vec k})$ at the interacting Fermi surface, and they do not count excited quasiparticles. Subtracting the two integrands,
\begin{equation}
 N_{1} - N_{2} = \intk \bigl[\Theta(\Phi_{\lambda\vec k}) - n_{\lambda\vec k}\bigr] = V_{F} - N_{\mathrm{tot}} ,
 \label{eq:S_countidentity}
\end{equation}
so that Luttinger's theorem is the statement $N_{1} = N_{2}$. By Eqs.~\eqref{eq:S_local} and \eqref{eq:S_Lsign} each population is a partial integral of Luttinger's integral,
\begin{equation}
 N_{1} = \intin \LL ,
 \qquad
 N_{2} = -\intout \LL ,
 \label{eq:S_NLparts}
\end{equation}
and therefore [Eqs.~\eqref{eq:antiflow_N12_is_L} and \eqref{eq:antiflow_L1norm} of the main text]
\begin{align}
N_{1-2}(\lambda) &\equiv V_{F} - N_{\mathrm{tot}} = N_{1} - N_{2} = \intk \LL_{\lambda\vec k} ,
 \label{eq:S_Delta} \\
N_{1+2}(\lambda) &\equiv N_{1} + N_{2} = \intk |\LL_{\lambda\vec k}| ,
 \label{eq:S_T} \\
 N_{1} &= \tfrac12(N_{1+2} + N_{1-2}) ,
 \qquad
 N_{2} = \tfrac12(N_{1+2} - N_{1-2}) .
 \label{eq:S_NDT}
\end{align}
 Luttinger's theorem, $N_{1-2} = 0$, is not assumed until Sec.~\ref{sec:S_LW}; everything before that section holds for any $N_{1-2}$.

\section{Coupling derivatives and the motion of the Fermi surface}
\label{sec:S_geometry}

Along the path, $\Sigma_{\lambda}(z,\vec k)$ is the self-energy of the ground state of $H_{\lambda}$ at the fixed density. Its derivative $\partial_{\lambda}\Sigma_{\lambda}(z,\vec k)$ at fixed $z$ and $\vec k$ contains both the explicit dependence on the interaction and the change of the ground state along the path (Sec.~\ref{sec:S_mudrift}). A frequency-dependent quantity $X_{\lambda}(\omega,\vec k)$ evaluated at the Fermi level changes in addition because $\mu$ moves. Its derivative along the path at fixed momentum is
\begin{equation}
 \D X_{\lambda}(\mu,\vec k) = \bigl[\partial_{\lambda}X_{\lambda}(\omega,\vec k) + \mu'\,\partial_{\omega}X_{\lambda}(\omega,\vec k)\bigr]_{\omega = \mu} ,
 \qquad
 \mu' = \frac{d\mu}{d\lambda} .
 \label{eq:S_Dfreq}
\end{equation}
The second term follows the drift of the Fermi level. Here and below a prime denotes the derivative $d/d\lambda$ along the path, as in $\mu'$, $V'_{F}$, and $N'_{\mathrm{tot}}$. Formally $\D$ is a convective derivative, with $\lambda$ as time and $\omega$ as position; it is still taken at fixed momentum. For a quantity $F_{\lambda\vec k}$ defined at each momentum and jumping at $\FS_{\lambda}$, such as $n_{\lambda\vec k}$ or $\LL_{\lambda\vec k}$, $\D F_{\lambda\vec k}$ is its rate of change at fixed $\vec k$ on either side of the surface; it does not include the motion of the jump, which is treated in Sec.~\ref{sec:S_transport}. Applied to Eq.~\eqref{eq:S_phi}, Eq.~\eqref{eq:S_Dfreq} gives
\begin{equation}
 \D\Phi_{\lambda\vec k} = \mu'\bigl[1 - \partial_{\omega}\Re\Sigma^{R}_{\lambda}(\mu,\vec k)\bigr] - \partial_{\lambda}\Re\Sigma^{R}_{\lambda}(\mu,\vec k) .
 \label{eq:S_Dphi}
\end{equation}
The function $\Phi$ changes through the self-energy at fixed frequency, through the frequency argument $\mu(\lambda)$ of the self-energy, and through the explicit $\mu$. On the Fermi surface the bracket is $Z^{-1}_{\lambda\vec k}$.

Consider now a point $\vec k_{\FS}(\lambda)$ that moves with the Fermi surface, $\Phi_{\lambda}(\vec k_{\FS}(\lambda)) = 0$, with velocity $\vec k'_{\FS} = d\vec k_{\FS}/d\lambda$. Differentiating this condition and projecting the velocity on the normal $\hat n = \nabla_{\vec k}\Phi/|\nabla_{\vec k}\Phi|$, which points into the sea, gives the normal speed of the surface,
\begin{equation}
 0 = \D\Phi_{\lambda\vec k} + \vec k'_{\FS}\cdot\nabla_{\vec k}\Phi_{\lambda\vec k} ,
 \qquad
 v_{\lambda\vec k} = \vec k'_{\FS}\cdot\hat n = -\frac{\D\Phi_{\lambda\vec k}}{|\nabla_{\vec k}\Phi_{\lambda\vec k}|} .
 \label{eq:S_vgeom}
\end{equation}
In our notations, a positive $v_{\lambda\vec k}$ means that the sea recedes. Only the normal component is fixed by the motion of the surface; a tangential component merely relabels its points. With Eqs.~\eqref{eq:S_gradient} and \eqref{eq:S_Dphi}, the normal speed takes the physical form [Eq.~\eqref{eq:hydro_vn_physical} of the main text]

\begin{equation}
 v_{\lambda\vec k} = \frac{Z_{\lambda\vec k}\,\partial_{\lambda}\Re\Sigma^{R}_{\lambda}(\mu,\vec k) - \mu'}{|\nabla_{\vec k}\tilde\omega_{\lambda\vec k}|} .
 \label{eq:S_vphysical}
\end{equation}
The numerator is the drift of the quasiparticle level relative to the chemical potential, since differentiating Eq.~\eqref{eq:S_qp} at fixed momentum gives

\begin{equation}
 \D\tilde\omega_{\lambda\vec k} = Z_{\lambda\vec k}\,\partial_{\lambda}\Re\Sigma^{R}_{\lambda}(\tilde\omega_{\lambda\vec k},\vec k) ,
 \label{eq:S_qpdrift}
\end{equation}
and the denominator is the magnitude of the quasiparticle velocity. The speed $v_{\lambda\vec k}$ has units of momentum per unit coupling: it describes how the Fermi surface deforms as the interaction is switched on, not a motion in real time.

A quantity $q_{\lambda}$ that lives on the Fermi surface, such as $Z_{\lambda\vec k}$ or one of the limits $n^{\mp}_{\lambda\vec k}$, is differentiated along the moving surface,
\begin{equation}
 \DS q = \frac{d}{d\lambda}\, q_{\lambda}\bigl(\vec k_{\FS}(\lambda)\bigr) .
 \label{eq:S_DS}
\end{equation}
If $q$ has a smooth continuation off the surface, this is
\begin{equation}
 \DS q = \D q + \vec k'_{\FS}\cdot\nabla_{\vec k}q .
 \label{eq:S_DSextension}
\end{equation}
For the limits of $n_{\lambda\vec k}$, whose slope normal to the surface may have an integrable singularity there, Eq.~\eqref{eq:S_DS} is the definition: only values on the surface enter it. Such singularities are not universal. They are absent for free fermions, for a static self-energy, and more generally for a self-energy that is real in a window around $\mu$, such as one with isolated real poles. 

A $\delta$ function of $\Phi$ turns a zone integral into an integral over the Fermi surface,
\begin{equation}
 \intk f(\vec k)\,\delta(\Phi_{\lambda\vec k}) = \ints \frac{f(\vec k)}{|\nabla_{\vec k}\Phi_{\lambda\vec k}|} .
 \label{eq:S_coarea}
\end{equation}
It is convenient to define the surface $\delta$ function
\begin{equation}
 \delta_{\FS_{\lambda}}(\vec k) = |\nabla_{\vec k}\Phi_{\lambda\vec k}|\,\delta(\Phi_{\lambda\vec k}) ,
 \qquad
 \intk f\,\delta_{\FS_{\lambda}} = \ints f .
 \label{eq:S_surfdelta}
\end{equation}
For a quantity that jumps at the moving surface, $d/d\lambda$ denotes its derivative at fixed $\vec k$ taken in the sense of distributions in $\vec k$;  the complete derivative, including the $\delta$ function produced by the motion of the jump, whereas $\D$ is the classical derivative on either side of the surface.   For the sharp Fermi sea itself, Eq.~\eqref{eq:S_vgeom} gives
\begin{equation}
 \dlam\Theta(\Phi_{\lambda\vec k}) = \delta(\Phi_{\lambda\vec k})\,\D\Phi_{\lambda\vec k} = -v_{\lambda\vec k}\,\delta_{\FS_{\lambda}} ,
 \qquad
 V'_{F} = -\ints v_{\lambda\vec k} .
 \label{eq:S_volumerate}
\end{equation}
A positive inward speed removes volume from the sea.

\section{Transport theorem for a moving discontinuity}
\label{sec:S_transport}

Let $F_{\lambda\vec k}$ change smoothly with $\lambda$ at fixed $\vec k$ on either side of the Fermi surface and jump across it, as $n_{\lambda\vec k}$ and $\LL_{\lambda\vec k}$ do. Near the surface we write it as
\begin{equation}
 F_{\lambda\vec k} = \Theta(\Phi_{\lambda\vec k})\,F^{\mathrm{in}}_{\lambda\vec k} + \Theta(-\Phi_{\lambda\vec k})\,F^{\mathrm{out}}_{\lambda\vec k} ,
 \label{eq:S_Fsplit}
\end{equation}
where, to begin with, $F^{\mathrm{in}}$ and $F^{\mathrm{out}}$ continue $F$ smoothly from inside and from outside across the surface. The derivative contains the rates on the two sides and the $\delta$ function of the moving step, whose coefficient is the jump,
\begin{align}
 \dlam F &= \Theta(\Phi)\,\D F^{\mathrm{in}} + \Theta(-\Phi)\,\D F^{\mathrm{out}} + (F^{-}_{\lambda\vec k} - F^{+}_{\lambda\vec k})\,\delta(\Phi)\,\D\Phi
 \notag \\
 &= \D F - [F_{\lambda\vec k}]\,v_{\lambda\vec k}\,\delta_{\FS_{\lambda}} .
 \label{eq:S_Fdis}
\end{align}
Only the values of the continuations on the surface, $F^{\mathrm{in}} = F^{-}_{\lambda\vec k}$ and $F^{\mathrm{out}} = F^{+}_{\lambda\vec k}$, enter the last term, so the result does not depend on how $F$ is continued. Integrating over the zone gives [Eq.~\eqref{eq:antiflow_jump} of the main text]
\begin{equation}
 \dlam \intk F_{\lambda\vec k} = \intk \D F_{\lambda\vec k} - \ints [F_{\lambda\vec k}]\,v_{\lambda\vec k} ,
 \label{eq:S_reynolds}
\end{equation}
the Reynolds transport theorem of fluid mechanics for a field with a moving discontinuity~\cite{Truesdell1960,Aris1962}.  The first term is the change of $F$ at fixed momentum on either side of the moving surface; the second is the jump carried along by the surface, $-v_{\lambda\vec k}\,dS_{\vec k}/(2\pi)^{d}$ being the volume gained by the sea through the element $dS_{\vec k}$.

A quantity counted only inside the sea, $\Theta(\Phi)F$, has the jump $F^{-}_{\lambda\vec k}$, because its value outside is zero; a quantity counted only outside, $\Theta(-\Phi)F$, has the jump $-F^{+}_{\lambda\vec k}$. Therefore [Eq.~\eqref{eq:flow_lift} of the main text, written there for $F_{\lambda\vec k} = n_{\lambda\vec k}$]
\begin{equation}
 \dlam\bigl[\Theta(\pm\Phi)F\bigr] = \Theta(\pm\Phi)\,\D F \pm \delta(\Phi)\,\D\Phi\, F^{\mp}_{\lambda\vec k}
 = \Theta(\pm\Phi)\,\D F \mp v_{\lambda\vec k}F^{\mp}_{\lambda\vec k}\,\delta_{\FS_{\lambda}} .
 \label{eq:S_stepweak}
\end{equation}
The surface term carries the value of $F$ on the side of the region, so that products such as $\Theta(\Phi)\,\delta(\Phi)$, which would require the value of a step at its own discontinuity, never appear. Integrated over the zone, the surface term is $\mp\ints F^{\mp}_{\lambda\vec k}v_{\lambda\vec k}$, which gives the transport formulas for the sea and for the rest of the zone,
\begin{align}
 \dlam \intin F &= \intin \D F - \ints F^{-}_{\lambda\vec k}v_{\lambda\vec k} ,
 \label{eq:S_insideReynolds} \\
 \dlam \intout F &= \intout \D F + \ints F^{+}_{\lambda\vec k}v_{\lambda\vec k} .
 \label{eq:S_outsideReynolds}
\end{align}
Neither formula requires the Fermi volume to be constant.

\section{Rates of the momentum distribution and of Luttinger's integral}
\label{sec:S_rates}

Away from the Fermi surface, the rates of $n_{\lambda\vec k}$ and $\LL_{\lambda\vec k}$ at fixed momentum follow from their frequency integrals. The upper limit $\mu(\lambda)$ moves, and the Leibniz rule adds the integrand at the Fermi level (the integrals are differentiated under the integral sign, as assumed in Sec.~\ref{sec:S_spectral}). For the occupation, Eq.~\eqref{eq:S_nzero} gives
\begin{equation}
 \D n_{\lambda\vec k} = \int_{-\infty}^{\mu} d\omega\, \partial_{\lambda}A_{\lambda}(\omega,\vec k) + \mu' A_{\lambda}(\mu,\vec k) .
 \label{eq:S_nendpoint}
\end{equation}
For Luttinger's integral, Eq.~\eqref{eq:S_Ldef} gives a different endpoint term,
\begin{equation}
 \D\LL_{\lambda\vec k} = \frac{1}{\pi}\Im\int_{-\infty}^{\mu} d\omega\, \partial_{\lambda}\bigl(\G^{R}_{\lambda}\,\partial_{\omega}\Sigma^{R}_{\lambda}\bigr)
 + \frac{\mu'}{\pi}\Im\bigl[\G^{R}_{\lambda}(\mu,\vec k)\,\partial_{\omega}\Sigma^{R}_{\lambda}(\mu,\vec k)\bigr] .
 \label{eq:S_Lendpoint}
\end{equation}
Because $\Im\Sigma^{R}_{\lambda} \le 0$ everywhere and $\Im\Sigma^{R}_{\lambda}(\mu,\vec k) = 0$, the damping has a minimum at the chemical potential, $\partial_{\omega}\Im\Sigma^{R}_{\lambda}(\mu,\vec k) = 0$, and the endpoint term of $\LL$ equals
\begin{equation}
 -\mu' A_{\lambda}(\mu,\vec k)\,\partial_{\omega}\Re\Sigma^{R}_{\lambda}(\mu,\vec k) .
 \label{eq:S_Lendpointreal}
\end{equation}
It differs from that of $n$ by the factor $-\partial_{\omega}\Re\Sigma^{R}_{\lambda}(\mu,\vec k)$. In the same way, differentiating $\Im\Sigma^{R}_{\lambda}(\mu(\lambda),\vec k) = 0$ along the path gives $\Im\partial_{\lambda}\Sigma^{R}_{\lambda}(\mu,\vec k) = 0$. Off the Fermi surface $\G^{R}_{\lambda}(\mu,\vec k) = 1/\Phi_{\lambda\vec k}$ is real and finite, and both endpoint terms vanish; on the surface they are singular, and we return to them at the end of this section.

Off the surface $\Theta(\Phi_{\lambda\vec k})$ does not change, and differentiating Luttinger's decomposition~\eqref{eq:S_local} at fixed momentum gives the first line of Eq.~\eqref{eq:antiflow_rates_Z} of the main text,
\begin{equation}
 \D n_{\lambda\vec k} = -\D\LL_{\lambda\vec k} ,
 \qquad \vec k \notin \FS_{\lambda} .
 \label{eq:S_bulkrate}
\end{equation}
This can be checked directly. Dyson's equation gives
\begin{equation}
 \partial_{\lambda}\G = \G^{2}\,\partial_{\lambda}\Sigma ,
 \qquad
 \partial_{\omega}\G = -\G^{2}(1 - \partial_{\omega}\Sigma) ,
 \label{eq:S_Gderivs}
\end{equation}
and hence
\begin{equation}
 \partial_{\lambda}(\G\,\partial_{\omega}\Sigma) = \G^{2}\,\partial_{\lambda}\Sigma + \partial_{\omega}(\G\,\partial_{\lambda}\Sigma) .
 \label{eq:S_Lderivintegrand}
\end{equation}
The total $\omega$-derivative integrates to its values at the two ends, of which the one at $\omega \to -\infty$ vanishes by Eq.~\eqref{eq:S_uvcondition}, and $\partial_{\lambda}A = -\pi^{-1}\Im[\G^{2}\partial_{\lambda}\Sigma]$ cancels the first term of Eq.~\eqref{eq:S_Lderivintegrand} under the integral. Adding Eqs.~\eqref{eq:S_nendpoint} and \eqref{eq:S_Lendpoint}, only terms at $\omega = \mu$ survive,
\begin{equation}
 \D(n + \LL) = \frac{1}{\pi}\Im\Bigl\{\G^{R}(\mu,\vec k)\bigl[\partial_{\lambda}\Sigma^{R}(\mu,\vec k) + \mu'\bigl(\partial_{\omega}\Sigma^{R}(\mu,\vec k) - 1\bigr)\bigr]\Bigr\} = 0
 \label{eq:S_bulkcheck}
\end{equation}
because every quantity in the braces is real off the surface. In a microscopic calculation the bulk rate is therefore
\begin{equation}
 \D n_{\lambda\vec k} = -\frac{1}{\pi}\Im\int_{-\infty}^{\mu} d\omega\, \bigl[\G^{R}_{\lambda}(\omega,\vec k)\bigr]^{2}\,\partial_{\lambda}\Sigma^{R}_{\lambda}(\omega,\vec k) ,
 \qquad \vec k \notin \FS_{\lambda} .
 \label{eq:S_bulkexplicit}
\end{equation}
This rate involves the full frequency dependence of $\G^{R}_{\lambda}$ and of $\partial_{\lambda}\Sigma^{R}_{\lambda}$; the residue at the Fermi surface does not determine it.

On the surface, Migdal's jump~\eqref{eq:S_migdal} and the jump~\eqref{eq:S_Ljump} of $\LL$, differentiated along the moving surface with Eq.~\eqref{eq:S_DS}, give the second line of Eq.~\eqref{eq:antiflow_rates_Z},

\begin{equation}
 \DS[n_{\lambda\vec k}] = -\DS[\LL_{\lambda\vec k}] = \DS Z_{\lambda\vec k} .
 \label{eq:S_surfacerate}
\end{equation}
The two lines involve different derivatives. The second cannot be replaced by the jump of the fixed-momentum rates: these rates may have integrable singularities at the surface (Sec.~\ref{sec:S_geometry}), so they need not have finite limits there; where $\D n$ and $\nabla_{\vec k}n$ have one-sided limits, following the surface still adds the motion of the observation point, $\DS[n_{\lambda\vec k}] = [\D n] + \vec k'_{\FS}\cdot[\nabla_{\vec k}n]$ by Eq.~\eqref{eq:S_DSextension}.

For $n$ and $\LL$, whose jumps are $Z_{\lambda\vec k}$ and $1 - Z_{\lambda\vec k}$, Eq.~\eqref{eq:S_Fdis} gives the complete derivatives

\begin{equation}
 \frac{dn_{\lambda\vec k}}{d\lambda} = \D n - Z_{\lambda\vec k}\,v_{\lambda\vec k}\,\delta_{\FS_{\lambda}} ,
 \qquad
 \frac{d\LL_{\lambda\vec k}}{d\lambda} = \D\LL - (1 - Z_{\lambda\vec k})\,v_{\lambda\vec k}\,\delta_{\FS_{\lambda}} .
 \label{eq:S_nLdis}
\end{equation}
Their sum is $-v_{\lambda\vec k}\delta_{\FS_{\lambda}} = d\Theta(\Phi_{\lambda\vec k})/d\lambda$, as Luttinger's decomposition~\eqref{eq:S_local} requires. The singular endpoint terms of Eqs.~\eqref{eq:S_nendpoint} and \eqref{eq:S_Lendpoint} on the surface belong to these swept-jump terms. Since $\Sigma^{R}_{\lambda}(\mu,\vec k)$ and $\partial_{\omega}\Sigma^{R}_{\lambda}(\mu,\vec k)$ are real, a common regulator $\eta \to 0^{+}$ gives $\G_{\lambda}(\mu + i\eta,\vec k)^{-1} = \Phi_{\lambda\vec k} + i\eta\,a_{\vec k} + o(\eta)$, with $a_{\vec k} = 1 - \partial_{\omega}\Re\Sigma^{R}_{\lambda}(\mu,\vec k)$ equal to $Z^{-1}_{\lambda\vec k} > 0$ on the surface; the width $\eta a_{\vec k}$ of the resulting Lorentzian in $\Phi$ does not change its weight, and as functions of $\vec k$

\begin{equation}
 A_{\lambda}(\mu,\vec k) = \delta(\Phi_{\lambda\vec k}) ,
 \qquad
 \frac{1}{\pi}\Im\bigl[\G^{R}_{\lambda}\,\partial_{\omega}\Sigma^{R}_{\lambda}\bigr](\mu,\vec k) = \frac{1 - Z_{\lambda\vec k}}{Z_{\lambda\vec k}}\,\delta(\Phi_{\lambda\vec k}) .
 \label{eq:S_endpointdelta}
\end{equation}
As distributions in $\vec k$, $\delta(\Phi_{\lambda\vec k}) = Z_{\lambda\vec k}\,\delta(\mu - \tilde\omega_{\lambda\vec k})$ by Eqs.~\eqref{eq:S_coarea} and \eqref{eq:S_gradient}, so the quasiparticle weight is already contained in $A_{\lambda}(\mu,\vec k) = \delta(\Phi_{\lambda\vec k})$. With $\D\Phi = \mu'/Z_{\lambda\vec k} - \partial_{\lambda}\Re\Sigma^{R}_{\lambda}(\mu,\vec k)$ on the surface, from Eq.~\eqref{eq:S_Dphi}, the swept-jump terms of Eq.~\eqref{eq:S_nLdis} are $Z_{\lambda\vec k}\,\delta(\Phi)\,\D\Phi$ and $(1 - Z_{\lambda\vec k})\,\delta(\Phi)\,\D\Phi$. Their parts proportional to $\mu'$ are $\mu'$ times the two expressions in Eq.~\eqref{eq:S_endpointdelta}, the Leibniz terms. The remaining parts, $-Z_{\lambda\vec k}\,\delta(\Phi)\,\partial_{\lambda}\Re\Sigma^{R}_{\lambda}$ and $-(1 - Z_{\lambda\vec k})\,\delta(\Phi)\,\partial_{\lambda}\Re\Sigma^{R}_{\lambda}$, are the singular parts of the fixed-frequency integrals in Eqs.~\eqref{eq:S_nendpoint} and \eqref{eq:S_Lendpoint}, which arise when the quasiparticle peak crosses $\omega = \mu$. Both are counted once, in the surface term; $\D n$ and $\D\LL$ denote the rates on either side of the surface.

\section{Flow equations for the deficit and the surplus}
\label{sec:S_flows}

The rates of $N_{1}$ and $N_{2}$ contain both the change of the occupation and the motion of the regions of integration [Eqs.~\eqref{eq:setup_dN1}--\eqref{eq:setup_dN2_theta} of the main text],
\begin{align}
 \frac{dN_{1}}{d\lambda} &= \dlam \intin (1 - n) = \dlam \intk \Theta(\Phi)\,(1 - n) ,
 \label{eq:S_N1derivdef} \\
 \frac{dN_{2}}{d\lambda} &= \dlam \intout n = \dlam \intk \Theta(-\Phi)\, n .
 \label{eq:S_N2derivdef}
\end{align}
The constant $1$ in the integrand of $N_{1}$ contributes only through the motion of the sea. The transport formulas~\eqref{eq:S_insideReynolds} and \eqref{eq:S_outsideReynolds} give [Eqs.~\eqref{eq:flow_dN1} and \eqref{eq:flow_dN2} of the main text]
\begin{align}
 \frac{dN_{1}}{d\lambda} &= -\ints (1 - n^{-}_{\lambda\vec k})\,v_{\lambda\vec k} - \intin \D n ,
 \label{eq:S_N1nrate} \\
 \frac{dN_{2}}{d\lambda} &= +\ints n^{+}_{\lambda\vec k}\,v_{\lambda\vec k} + \intout \D n .
 \label{eq:S_N2nrate}
\end{align}
The signs are fixed by geometry. Where the surface recedes, $v_{\lambda\vec k} > 0$, momenta with hole occupation $1 - n^{-}_{\lambda\vec k}$ leave the sea and are no longer counted in $N_{1}$, while the exterior gains momenta with occupation $n^{+}_{\lambda\vec k}$, which are counted in $N_{2}$. With Eq.~\eqref{eq:S_traceweights}, which follows from Migdal's jump, in the surface term and the bulk identity~\eqref{eq:S_bulkrate} in the bulk terms, these become the flow equations [Eqs.~\eqref{eq:hydro_flows_equal_N1} and \eqref{eq:hydro_flows_equal_N2} of the main text],

\begin{align}
 \frac{dN_{1}}{d\lambda} &= -\ints (1 - Z_{\lambda\vec k} - n^{+}_{\lambda\vec k})\,v_{\lambda\vec k} + \intin \D\LL ,
 \label{eq:S_flow1} \\
 \frac{dN_{2}}{d\lambda} &= +\ints n^{+}_{\lambda\vec k}\,v_{\lambda\vec k} - \intout \D\LL .
 \label{eq:S_flow2}
\end{align}
Neither a Luttinger--Ward functional nor $N_{1-2} = 0$ has been used. The equations involve the motion of the Fermi surface, the occupations just inside and just outside it, and the bulk rate of Luttinger's integral, which depends on the full spectral function.

The two surface terms are not equal. The first can be split as

\begin{equation}
 -\ints (1 - Z_{\lambda\vec k} - n^{+}_{\lambda\vec k})\,v_{\lambda\vec k} = -\ints (1 - Z_{\lambda\vec k})\,v_{\lambda\vec k} + \ints n^{+}_{\lambda\vec k}\,v_{\lambda\vec k} .
 \label{eq:S_sharedflux}
\end{equation}
The flux of the surplus occupation, $\ints n^{+}_{\lambda\vec k}v_{\lambda\vec k}$, is common to both equations and cancels in their difference, while the flux of the incoherent weight $1 - Z_{\lambda\vec k}$ enters only the rate of $N_{1}$. The difference and the sum of the flow equations are

\begin{align}
 \dlam (N_{1} - N_{2}) &= \intk \D\LL - \ints (1 - Z_{\lambda\vec k})\,v_{\lambda\vec k} ,
 \label{eq:S_differenceflow} \\
 \dlam (N_{1} + N_{2}) &= \intin \D\LL - \intout \D\LL - \ints (1 - Z_{\lambda\vec k} - 2n^{+}_{\lambda\vec k})\,v_{\lambda\vec k} .
 \label{eq:S_sumflow}
\end{align}
The same two equations follow directly from the transport theorem~\eqref{eq:S_reynolds} applied to $\LL$ and to $|\LL|$: by Eq.~\eqref{eq:S_Lsign}, $\D|\LL| = \D\LL$ inside the sea and $-\D\LL$ outside, and the surface terms are the jumps~\eqref{eq:S_Ljump} and \eqref{eq:S_absjump}. The flow equations are thus consistent with both counting identities, Eqs.~\eqref{eq:S_Delta} and \eqref{eq:S_T}.

\section{Particle-number conservation and surface--bulk balances}
\label{sec:S_balance}

The transport theorem applied to the occupation, with $[n_{\lambda\vec k}] = Z_{\lambda\vec k}$, gives

\begin{equation}
 N'_{\mathrm{tot}} = \intk \D n - \ints Z_{\lambda\vec k}\,v_{\lambda\vec k} .
 \label{eq:S_numbergeneral}
\end{equation}
At fixed density this is Eq.~\eqref{eq:hydro_number_flux} of the main text,
\begin{equation}
 \intk \D n_{\lambda\vec k} = \ints Z_{\lambda\vec k}\,v_{\lambda\vec k} .
 \label{eq:S_numberfixed}
\end{equation}
The corresponding statement for $\LL$ follows from Eq.~\eqref{eq:S_Delta}, or from Eq.~\eqref{eq:S_differenceflow},
\begin{equation}
 N'_{1-2} = \intk \D\LL - \ints (1 - Z_{\lambda\vec k})\,v_{\lambda\vec k} .
 \label{eq:S_Deltarate}
\end{equation}
At fixed density $N'_{1-2} = V'_{F} = -\ints v_{\lambda\vec k}$ by Eq.~\eqref{eq:S_volumerate}, and therefore

\begin{equation}
 \intk \D\LL = N'_{1-2} + \ints (1 - Z_{\lambda\vec k})\,v_{\lambda\vec k} = -\ints Z_{\lambda\vec k}\,v_{\lambda\vec k} ,
 \label{eq:S_Lbalancegeneral}
\end{equation}
in agreement with Eqs.~\eqref{eq:S_bulkrate} and \eqref{eq:S_numberfixed}. The middle equation reduces to $\ints (1 - Z_{\lambda\vec k})\,v_{\lambda\vec k}$ only when the Fermi volume is constant (Sec.~\ref{sec:S_mudrift}); in general the term $N'_{1-2}$ must be kept.

\section{Luttinger's theorem and its consequences}
\label{sec:S_LW}

\subsection{The Luttinger--Ward argument}
\label{sec:S_LWargument}

The kinematic results leave $N_{1-2}$ undetermined. Luttinger and Ward derived $N_{1-2} = 0$ from the existence of a functional $\mathscr{W}_{\lambda}[\G]$ of the Green's function whose functional derivative is the self-energy~\cite{LuttingerWard1960,Luttinger1960}, the functional that also generates conserving approximations~\cite{BaymKadanoff1961,Baym1962}. The existence of the functional, with $\Sigma_{\lambda} = \delta\mathscr{W}_{\lambda}/\delta\G_{\lambda}$, does not by itself establish the theorem. The derivation needs three further assumptions, stated where they enter below: that the invariance of the functional under a uniform shift of the frequencies survives the summation of the diagrams and the zero-temperature limit; that $\Sigma_{\lambda}$ is the derivative of the functional on its physical branch, since the map from $\G$ to $\Sigma$ can have several branches and the skeleton series can converge to an unphysical one; and that the limits and integrals can be taken in the order used, with no contribution concentrated on the Fermi surface. We briefly remind the argument to show where these assumptions enter. At zero temperature the Matsubara frequencies become continuous; we write $z = \mu + i\nu$ with real $\nu$. The Luttinger--Ward functional $\mathscr{W}_{\lambda}[\G]$ is the sum of the closed skeleton diagrams, and its variation gives the self-energy,
\begin{equation}
 \delta\mathscr{W}_{\lambda} = \intk \int_{-\infty}^{\infty} \frac{d\nu}{2\pi}\, \Sigma_{\lambda}(\mu + i\nu,\vec k)\,\delta\G_{\lambda}(\mu + i\nu,\vec k) .
 \label{eq:S_LWvariation}
\end{equation}
For number-conserving instantaneous interactions each skeleton diagram of $\mathscr{W}_{\lambda}$ is unchanged when the frequencies of all internal lines are shifted by the same amount, because the shifts cancel at every vertex. We assume that this invariance survives the summation of the diagrams and the zero-temperature limit, and that $\Sigma_{\lambda}$ is the derivative of the functional on its physical branch~\cite{KozikFerreroGeorges2015}. The shift $\G(z,\vec k) \to \G(z + i\alpha,\vec k)$ then gives
\begin{equation}
 0 = \frac{d\mathscr{W}_{\lambda}}{d\alpha}\Big|_{\alpha = 0}
 = i\intk \int_{-\infty}^{\infty} \frac{d\nu}{2\pi}\, \Sigma_{\lambda}(z,\vec k)\,\partial_{z}\G_{\lambda}(z,\vec k) .
 \label{eq:S_LWshift}
\end{equation}
An integration by parts in $\nu$, with $\partial_{\nu} = i\partial_{z}$ and no contribution from $|\nu| \to \infty$, then gives
\begin{equation}
 \intk \int_{-\infty}^{\infty} \frac{d\nu}{2\pi}\, \G_{\lambda}(z,\vec k)\,\partial_{z}\Sigma_{\lambda}(z,\vec k) = 0 .
 \label{eq:S_LWimag}
\end{equation}
The frequency integral is done at each momentum before the momentum integral.

To return to real frequencies, let $h(z)$ be analytic off the real axis, with $h(z^{*}) = h(z)^{*}$, and fall off faster than $1/|z|$; in addition, we take $h$ regular at $z = \mu$, where the line of integration crosses the real axis, that is, bounded near $z = \mu$ with finite boundary values $h(\mu \pm i0^{+})$. At momenta off the Fermi surface $\G\,\partial_{z}\Sigma$ is an example: there $\G_{\lambda}(\mu,\vec k) = 1/\Phi_{\lambda\vec k}$ is finite, and $\partial_{z}\Sigma_{\lambda}$ is finite and real at $z = \mu$. The integration runs along the line $\Re\,z = \mu$; closing the contour in the half-plane $\Re\,z < \mu$ and shrinking it onto the part of the real axis below $\mu$ gives
\begin{equation}
 \int_{-\infty}^{\infty} \frac{d\nu}{2\pi}\, h(\mu + i\nu) = -\frac{1}{\pi}\Im\int_{-\infty}^{\mu} d\omega\, h(\omega + i0^{+}) .
 \label{eq:S_contour}
\end{equation}
 Applied to $h = \G\,\partial_{z}\Sigma$ at momenta off the Fermi surface and integrated over the zone, with no further singular contributions, Eq.~\eqref{eq:S_LWimag} becomes Eq.~\eqref{eq:LW_real} of the main text,
\begin{equation}
 \intk \LL_{\lambda\vec k} = 0 .
 \label{eq:S_LWreal}
\end{equation}
With Eq.~\eqref{eq:S_Delta} this is Luttinger's theorem,
\begin{equation}
 V_{F} = N_{\mathrm{tot}} ,
 \qquad
 N_{1} = N_{2} .
 \label{eq:S_Luttinger}
\end{equation}
Once Eq.~\eqref{eq:S_LWreal} holds, by this or any other argument, the consequences below follow without further reference to the Luttinger-Ward functional.

\subsection{Equal rates of the deficit and the surplus}
\label{sec:S_equalrates}

If Eq.~\eqref{eq:S_LWreal} holds at every coupling on the path, Eqs.~\eqref{eq:S_T} and \eqref{eq:S_NDT} give Eq.~\eqref{eq:hydro_flows_L1} of the main text,
\begin{align}
 N_{1} = N_{2} = \frac12\intk |\LL_{\lambda\vec k}| ,
\\
 \frac{dN_{1}}{d\lambda} = \frac{dN_{2}}{d\lambda} = \frac12\dlam\intk |\LL_{\lambda\vec k}| .
 \label{eq:S_equalrates}
\end{align}
The derivative of the integral on the right contains the surface term of Eq.~\eqref{eq:S_sumflow}. Equality of $N_{1}$ and $N_{2}$ at the two ends of the path alone would not make their rates equal in between. Conversely, equal rates throughout the path fix $N_{1} - N_{2}$ only up to a constant, which vanishes at $\lambda = 0$, where both populations are zero.

\subsection{The surface--bulk balance}
\label{sec:S_incoherent}

With Eq.~\eqref{eq:S_LWreal} valid along the path, $N'_{1-2} = 0$, and Eq.~\eqref{eq:S_Deltarate} becomes Eq.~\eqref{eq:antiflow_zero} of the main text,

\begin{equation}
 \intk \D\LL_{\lambda\vec k} = \ints (1 - Z_{\lambda\vec k})\,v_{\lambda\vec k} .
 \label{eq:S_incoherentbalance}
\end{equation}
The Fermi surface sweeps a flux of incoherent weight $1 - Z_{\lambda\vec k}$, which the bulk change of Luttinger's integral compensates. Because $\ints v_{\lambda\vec k} = 0$, the right side also equals $-\ints Z_{\lambda\vec k}\,v_{\lambda\vec k}$, and with the bulk identity~\eqref{eq:S_bulkrate} the balance becomes the number balance~\eqref{eq:S_numberfixed}. The two balances are consequences of number conservation, Migdal's jump, and the constant Fermi volume. The two sides of Eq.~\eqref{eq:S_incoherentbalance} are in general non-zero.  It is the rate of $\int\LL$, the moving jump included, that vanishes.

\subsection{Constant Fermi volume and the drift of the chemical potential}
\label{sec:S_mudrift}

At fixed density Eq.~\eqref{eq:S_Luttinger} gives $V'_{F} = 0$, and Eq.~\eqref{eq:S_volumerate} becomes
\begin{equation}
 \ints v_{\lambda\vec k} = 0 ,
 \label{eq:S_zeroflux}
\end{equation}
that is, the net normal flux through the Fermi surface vanishes, while parts of the surface may advance and others recede. To solve for $\mu'$ we introduce the quasiparticle density of states on the Fermi surface,
\begin{equation}
 d\rho_{\vec k} = \frac{dS_{\vec k}}{(2\pi)^{d}\,|\nabla_{\vec k}\tilde\omega_{\lambda\vec k}|} ,
 \qquad
 \rho_{*} = \oint_{\FS_{\lambda}} d\rho_{\vec k} ,
 \qquad
 \langle q\rangle_{\FS_{\lambda}} = \frac{1}{\rho_{*}}\oint_{\FS_{\lambda}} q\, d\rho_{\vec k} ,
 \label{eq:S_DOSmeasure}
\end{equation}
with $0 < \rho_{*} < \infty$ for a regular Fermi surface away from a van Hove singularity; $\langle q\rangle_{\FS_{\lambda}}$ is the average over the surface weighted with the density of states. Inserting Eq.~\eqref{eq:S_vphysical} into Eq.~\eqref{eq:S_zeroflux} gives

\begin{equation}
 0 = \oint_{\FS_{\lambda}} d\rho_{\vec k}\,\bigl[Z_{\lambda\vec k}\,\partial_{\lambda}\Re\Sigma^{R}_{\lambda}(\mu,\vec k) - \mu'\bigr] ,
 \label{eq:S_muaverageintermediate}
\end{equation}
and solving for the drift of the chemical potential gives Eq.~\eqref{eq:hydro_mudot} of the main text,

\begin{equation}
 \mu' = \bigl\langle Z_{\lambda\vec k}\,\partial_{\lambda}\Re\Sigma^{R}_{\lambda}(\mu,\vec k)\bigr\rangle_{\FS_{\lambda}} .
 \label{eq:S_mudrift}
\end{equation}
The chemical potential follows the average drift of the quasiparticle levels on the Fermi surface. No further factor $Z_{\lambda\vec k}$ appears in the weight of the average: the conversion from $|\nabla_{\vec k}\Phi|$ to $|\nabla_{\vec k}\tilde\omega|$ has already produced the factor $Z_{\lambda\vec k}$ that multiplies $\partial_{\lambda}\Sigma$ in Eq.~\eqref{eq:S_vphysical}. Written with the measure $d\tilde\rho_{\vec k} = dS_{\vec k}/[(2\pi)^{d}\,|\nabla_{\vec k}\Phi_{\lambda\vec k}|] = Z_{\lambda\vec k}\,d\rho_{\vec k}$ of Eqs.~\eqref{eq:S_coarea} and \eqref{eq:S_gradient}, the same relation reads $\mu' = \oint_{\FS_{\lambda}} d\tilde\rho_{\vec k}\,\partial_{\lambda}\Re\Sigma^{R}_{\lambda}(\mu,\vec k) \big/ \oint_{\FS_{\lambda}} d\tilde\rho_{\vec k}\,Z^{-1}_{\lambda\vec k}$.

\section{Multiband systems}

We work at $T=0$ in a fixed, $\lambda$-independent orthonormal Bloch basis of $M$ orbital--spin
states per cell, with cell volume and $\hbar$ set to one. Hats denote matrices and $\id$ the identity;
$\Tr$ is the trace in this basis. With fixed bare Hamiltonian $\hat h_0(\kk)$
and self-energy $\SSig(z)$, the Green function and occupation matrix obey
\begin{subequations}
  \label{eq:green-definitions}
  \begin{equation}
    \GG^{-1}(z)=z\id-\hat h_0(\kk)-\SSig(z),\qquad
    \nn=\int_{-\infty}^{\mu_\lambda}d\omega\,\hat A_{\lambda\kk}(\omega).
    \label{eq:matrix}
  \end{equation}
  Here $\lambda$ is the interaction strength, $\mu_\lambda$ the chemical potential,
  and $z$ a complex frequency measured from a fixed energy zero.
  Matsubara frequencies correspond to $z=\mu_\lambda+i\omega_n$.
  For retarded ($R$) and advanced ($A$) functions,
  \begin{equation}
    \begin{gathered}
      \GG^{R/A}(\omega)=\lim_{\eta\to0^+}\GG(\omega\pm i\eta),\qquad
      \GG^A(\omega)=[\GG^R(\omega)]^\dagger,\\
      \hat A_{\lambda\kk}(\omega)=\frac{\GG^A(\omega)-\GG^R(\omega)}{2\pi i},\qquad
      (\nn)_{ab}=\langle c^\dagger_{\kk b}c_{\kk a}\rangle_\lambda.
    \end{gathered}
    \label{eq:spectral-definitions}
  \end{equation}
\end{subequations}
The self-energy $\SSig^{R/A}(\omega)$ obeys the same retarded/advanced relations.
Positive spectral weight and $\int d\omega\,\hat A_{\lambda\kk}(\omega)=\id$ give
$0\leq\nn\leq\id$. We assume a causal, number-conserving normal state with finite Hermitian
$\SSig^R(\mu_\lambda)$. All integrals are assumed to converge and permit differentiation.

The reference Hamiltonian $\hat h_0(\kk)+\SSig^R(\mu_\lambda)$ has orthonormal eigenstates
$|\unu\rangle$ with energies $\mu_\lambda+\xinu$:
\begin{equation}
    \bigl[\hat h_0(\kk)+\SSig^R(\mu_\lambda)-\mu_\lambda\id\bigr]|\unu\rangle
    =\xinu|\unu\rangle,
    \end{equation}
For the inverse Green's fucntion, defining interacting Fermi sheets, we thus have 
    \begin{equation}
    \Ph=\GG^{-1}(\mu_\lambda+i0^+)
    =\mu_\lambda\id-\hat h_0(\kk)-\SSig^R(\mu_\lambda)
    =-\sum_\nu\xinu|\unu\rangle\langle\unu|.
  \label{eq:phi}
\end{equation}
The Fermi sheets $S_{\lambda\nu}$ satisfy $\xinu=0$. The matrix step function selects the filled subspace:
\begin{equation}
  \Theta(\Ph)=\sum_{\nu:\,\xinu<0}|\unu\rangle\langle\unu|,
  \qquad \Theta(\Ph)[\id-\Theta(\Ph)]=0.
  \label{eq:projectors}
\end{equation}
$\id-\Theta(\Ph)$ selects the empty reference states.
Degeneracies within either subspace leave its projector well defined; $\nn$ and $\Theta(\Ph)$
need not commute. The counts per cell are
\begin{subequations}
  \label{eq:counts}
  \begin{align}
    N_1(\lambda)&=\ik\Tr[\Theta(\Ph)(\id-\nn)],
    \label{eq:hole-count} \\
    N_2(\lambda)&=\ik\Tr([\id-\Theta(\Ph)]\nn),
    \label{eq:particle-count} \\
    V_F(\lambda)&=\ik\Tr\Theta(\Ph),
    \label{eq:reference-count} \\
    N_{\rm tot}(\lambda)&=\ik\Tr\nn.
    \label{eq:total-count}
  \end{align}
\end{subequations}
$\nnu=\langle\unu|\nn|\unu\rangle$ is the mean occupation of a reference state.
$N_1(\lambda)$ counts ground-state holes $1-\nnu$ in filled states;
$N_2(\lambda)$ counts particles $\nnu$ in empty states.
$V_F(\lambda)$ includes filled bands; $N_{\rm tot}(\lambda)$ is the density.

\paragraph{Hole and particle weights.}
Off the Fermi sheets, tracing Dyson's equation and integrating over occupied frequencies gives (see also Ref. \cite{SekiYunoki2017}), 

  \begin{align}
    \Tr\GG(z)&=\partial_z\Tr\log\GG^{-1}(z)
    +\Tr[\GG(z)(\partial_z\SSig)(z)],
    \label{eq:trace-dyson} 
    \end{align} 
    hence
    \begin{align}
    \Ll&=\Tr[\Theta(\Ph)-\nn]
    =\frac{1}{\pi}\int_{-\infty}^{\mu_\lambda}d\omega\,
    \operatorname{Im}\Tr[\GG^R(\omega)(\partial_\omega\SSig^R)(\omega)].
    \label{eq:occupation-deficit}
  \end{align}
Applying $-\pi^{-1}\operatorname{Im}\int_{-\infty}^{\mu_\lambda}d\omega$
to Eq.~\eqref{eq:trace-dyson} at $z=\omega+i0^+$ gives $\Tr\nn$ on the left
by Eq.~\eqref{eq:spectral-definitions}. The trace-log boundary term gives
$\Tr\Theta(\Ph)$ from its phase change, see also Ref. \cite{SekiYunoki2017}. 

Hole and particle contributions can cancel in $\Ll$. Their separate weights are retained in the matrix
  \begin{align}
    \Lam=\Theta(\Ph)(\id-\nn)\Theta(\Ph)-[\id-\Theta(\Ph)]\nn[\id-\Theta(\Ph)],
      \label{eq:Lambda}
    \end{align}
 note that    
    \begin{align}
    |\Lam|=[2\Theta(\Ph)-\id]\Lam.
  \end{align}
Since $0\leq\nn\leq\id$, the filled block of $\Lam$ is nonnegative (hole weight) and
the empty block nonpositive (minus particle weight).
Thus $|\Lam|\equiv(\Lam^2)^{1/2}$ adds both weights: $2\Theta(\Ph)-\id$ reverses the empty block's sign.
$\Lam$ commutes with $\Theta(\Ph)$ even when $\nn$ does not.
Cyclicity and $\Theta(\Ph)^2=\Theta(\Ph)$ remove repeated projectors under the trace, giving
  \begin{align}
    \Ll&=\Tr\Lam=\Tr[\Theta(\Ph)-\nn],
    \end{align}
    and
    \begin{align}
    N_1(\lambda)-N_2(\lambda)&= \ik \Tr\Lam ,\\
    N_1(\lambda)+N_2(\lambda)&=\ik\Tr|\Lam|.
   \label{eq:deficit}
   \end{align}

\paragraph{Fermi-sheet motion.}
We assume each sheet $S_{\lambda\nu}$ has one nondegenerate crossing band, $\xinu=0$, with
$|\nabla_{\kk}\xinu|\ne0$, and that $(\partial_\omega\SSig^R)(\mu_\lambda)$ is finite and Hermitian.
On the sheet, the small-energy expansion is
  \begin{equation}
    \GG^{-1}(\mu_\lambda+\varepsilon+i0^+)
    =\Ph+\varepsilon[\id-(\partial_\omega\SSig^R)(\mu_\lambda)]+o(\varepsilon).
    \label{eq:pole-expansion}
  \end{equation}
  We therefore define the quasiparticle poles associated with each band
  \begin{align}
    \Znu^{-1} & =\langle\unu|\id-(\partial_\omega\SSig^R)(\mu_\lambda)|\unu\rangle,
    \label{eq:jump}
    \\
    \nn^- -\nn^+ & =\Znu|\unu\rangle\langle\unu|.
    \label{eq:jump2}
  \end{align}
We note that $0<\Znu\leq1$. Superscripts $-$ and $+$ in \label{eq:jump2}   denote limits from $\xinu<0$ and $\xinu>0$, giving 
$\nnu^- -\nnu^+=\Znu$.
On $S_{\lambda\nu}$, the quasiparticle velocity and inward normal momentum displacement per unit $\lambda$ are
\begin{equation}
  \begin{aligned}
   v^F_{\lambda \nu \vec k}  &=\Znu \nabla_{\kk}\xinu,\\
    \vnu&=\frac{\Dl\xinu}{|\nabla_{\kk}\xinu|}
    =\frac{\Znu\langle\unu|(\partial_\lambda\SSig^R)(\mu_\lambda)|\unu\rangle-\frac{d\mu_\lambda}{d\lambda}}
    {|v^F_{\lambda \nu \vec k}|}.
  \end{aligned}
  \label{eq:speed}
\end{equation}
$\vnu>0$ means the filled Fermi pocket shrinks.
$(\partial_\lambda\SSig^R)(\mu_\lambda)$ is taken at fixed frequency and includes the ground-state response.
Equation~\eqref{eq:speed} includes the chain-rule term
$\frac{d\mu_\lambda}{d\lambda}(\partial_\omega\SSig^R)(\mu_\lambda)$ from the moving chemical potential.

Away from the Fermi sheets, the number of filled reference states is fixed, $\Tr\Dl\Theta(\Ph)=0$,
but their orbital character can change. Both count integrands contain terms like $\Tr[\Theta(\Ph)\nn]$. We can further take its derivative
\begin{equation}
  \Dl\Tr[\Theta(\Ph)\nn]
  =\Tr[\Theta(\Ph)\Dl\nn]
  +\Tr\!\left\{[\Dl\Theta(\Ph)]\nn\right\}.
  \label{eq:rotation}
\end{equation}
The first term describes changing occupations; the second, rotation of the reference states relative to $\nn$.
Differentiating $\Theta(\Ph)^2=\Theta(\Ph)$ shows that $\Dl\Theta(\Ph)$ connects only filled and empty states.
Their off-diagonal occupation elements can therefore affect the rates, although they do not enter the counts.

\paragraph{Hole and particle flows.}
Crossing from $\xinu<0$ to $\xinu>0$ removes $|\unu\rangle\langle\unu|$ from $\Theta(\Ph)$.
Using Eq.~\eqref{eq:jump}, the hole and particle integrands jump by
$1-\nnu^-$ and $-\nnu^+$ (inside minus outside); the other bands cancel.
The surface contribution is minus the jump times the signed volume swept through momentum space.
Together with the response at fixed $\kk$, this gives
\begin{equation}
  \boxed{
  \frac{dN_1(\lambda)}{d\lambda}
  =-\sum_\nu\isnu(1-\nnu^-)\vnu
  -\ik\Tr\,\Dl\!\left[\Theta(\Ph)\nn\right].}
  \label{eq:flow1}
\end{equation}
\begin{equation}
  \boxed{
  \frac{dN_2(\lambda)}{d\lambda}
  =+\sum_\nu\isnu\nnu^+\vnu
  +\ik\Tr\,\Dl\!\left\{[\id-\Theta(\Ph)]\nn\right\}.}
  \label{eq:flow2}
\end{equation}

If we apply the transport theorem~\eqref{eq:S_reynolds} to the counts
defined in Eqs.~\eqref{eq:reference-count}, \eqref{eq:total-count}, using their respective integrand jumps
$1$, $Z_{\lambda\nu\mathbf{k}}$, and $1-Z_{\lambda\nu\mathbf{k}}$,
as follows from Eqs.~\eqref{eq:projectors}, \eqref{eq:jump},
and \eqref{eq:occupation-deficit}, we obtain

\begin{align}
    \frac{dV_F(\lambda)}{d\lambda}=-\sum_\nu\isnu\vnu,
\ \ \ 
    \frac{dN_{\rm tot}(\lambda)}{d\lambda}=\ik\Tr\Dl\nn-\sum_\nu\isnu\Znu\vnu,\\
  \label{eq:balances}
\end{align}
and for the difference
\begin{align}
    \frac{dN_{1-2}(\lambda)}{d\lambda}
    =\frac{d}{d\lambda}\ik\Tr\Lam
    =\ik\Dl\Ll-\sum_\nu\isnu(1-\Znu)\vnu.
  \label{eq:balances}
\end{align}
These identities do not require Luttinger's theorem. The full derivative, as previously, includes sheet motion. 
At fixed density, the bulk occupation change balances the moving jumps.

If we further assume $V_F(\lambda)=N_{\rm tot}(\lambda)$ (Luttinger theorem) holds along the path, then we obtain 
\begin{align}
  \frac{dN_1(\lambda)}{d\lambda}=\frac{dN_2(\lambda)}{d\lambda}=\frac12\frac{d}{d\lambda}\ik\Tr|\Lam|,
  \\
  \ik\Dl\Ll=\sum_\nu\isnu(1-\Znu)\vnu.
  \label{eq:conditional}
\end{align}
Note that fixed density the total reference volume is then constant, but individual sheets may still move.

\

\paragraph{Single-FS reduction.}
Consider now the single-band Fermi surface. For $M=1$, orbital rotation is absent. Omitting hats and $\nu$, off the Fermi surface we have
$
  \xi_{\lambda\kk}=-\Phi_{\lambda\kk}$, 
  $\Dl\Theta(\Phi_{\lambda\kk})=0$, 
 $ \Lambda_{\lambda\kk}=\Ll=\Theta(\Phi_{\lambda\kk})-n_{\lambda\kk}$.
Using $|\Lambda_{\lambda\kk}|=|\Ll|$, $\Dl n_{\lambda\kk}=-\Dl\Ll$, and
$n_{\lambda\kk}^-=Z_{\lambda\kk}+n_{\lambda\kk}^+$, Eqs.~\eqref{eq:flow1}--\eqref{eq:flow2}
reduce to main-text Eqs.~(3)--(4):
\begin{equation}
  \begin{aligned}
    \frac{dN_1(\lambda)}{d\lambda}
    &=-\oint_{S_\lambda}\frac{dS_{\kk}}{(2\pi)^d}
    (1-Z_{\lambda\kk}-n_{\lambda\kk}^+)v_{\lambda\kk}
    +\int_{\Phi_{\lambda\kk}>0}\frac{d^d\kk}{(2\pi)^d}\Dl\Ll,\\[3pt]
    \frac{dN_2(\lambda)}{d\lambda}
    &=+\oint_{S_\lambda}\frac{dS_{\kk}}{(2\pi)^d}
    n_{\lambda\kk}^+v_{\lambda\kk}
    -\int_{\Phi_{\lambda\kk}<0}\frac{d^d\kk}{(2\pi)^d}\Dl\Ll.
  \end{aligned}
  \label{eq:scalar-flows}
\end{equation}
$S_\lambda$ includes all components. Eq.~\eqref{eq:speed} becomes main-text Eq.~(23),
with  quasiparticle energy $\widetilde\omega_{\lambda\kk}$,
\begin{equation}
  v_{\lambda\kk}
  =-\frac{\Dl\Phi_{\lambda\kk}}{|\nabla_{\kk}\Phi_{\lambda\kk}|}
  =\frac{Z_{\lambda\kk}(\partial_\lambda\Sigma_{\lambda\kk}^R)(\mu_\lambda)-\frac{d\mu_\lambda}{d\lambda}}
  {|\nabla_{\kk}\widetilde\omega_{\lambda\kk}|}.
  \label{eq:scalar-speed}
\end{equation}
\end{document}